\documentclass[twocolumn,amsmath,amssymb,floatfix,superscriptaddress,prl]{revtex4-2}
\usepackage{url,epsf}
\usepackage{comment}
\usepackage{appendix}
\usepackage{physics}
\usepackage{multirow}
\usepackage[colorlinks=true,linkcolor=blue]{hyperref}
\usepackage{amsmath,bm,graphicx,epsfig}
\usepackage{xcolor}
\usepackage[english]{babel}
\usepackage{amsfonts,stmaryrd,amssymb}
\usepackage{enumerate}

\newcommand{\bQ}{\mathbf{Q}}

\newcommand{\bA}{\mathbf{A}}

\newcommand{\bq}{\mathbf{q}}
\newcommand{\bk}{\mathbf{k}}
\newcommand{\bp}{\mathbf{p}}

\newcommand{\br}{\mathbf{r}}
\newcommand{\eqn}[1]{(\ref{#1})}
\newcommand{\mr}{moir\'e~}

\begin{document}
\title{Chiral Kondo Lattice in Doped MoTe$_2$/WSe$_2$ Bilayers}

\author{Daniele Guerci}
\affiliation{Center for Computational Quantum Physics, Flatiron Institute, New York, New York 10010, USA}

\author{Jie Wang}
\affiliation{Center for Computational Quantum Physics, Flatiron Institute, New York, New York 10010, USA}

\author{Jiawei Zang}
\affiliation{Department of Physics, Columbia University, 538 West 120th Street, New York, NY 10027}

\author{Jennifer Cano}
\affiliation{Department of Physics and Astronomy, Stony Brook University, Stony Brook, New York 11794, USA}
\affiliation{Center for Computational Quantum Physics, Flatiron Institute, New York, New York 10010, USA}

\author{J. H. Pixley}
\affiliation{Department of Physics and Astronomy, Center for Materials Theory, Rutgers University, Piscataway, New Jersey 08854, USA}
\affiliation{Center for Computational Quantum Physics, Flatiron Institute, New York, New York 10010, USA}

\author{Andrew Millis}
\affiliation{Center for Computational Quantum Physics, Flatiron Institute, New York, New York 10010, USA}
\affiliation{Department of Physics, Columbia University, 538 West 120th Street, New York, NY 10027}

\begin{abstract}
We theoretically study the interplay between magnetism and a heavy Fermi liquid in the AB stacked transition metal dichalcogenide bilayer system MoTe$_2$/WSe$_2$ in the regime in which the $Mo$ layer supports localized magnetic moments coupled by interlayer electron tunnelling to a weakly correlated band of itinerant electrons in the $W$ layer. 
We show that the interlayer electron transfer leads to a chiral Kondo exchange, with consequences including a strong dependence of the Kondo temperature on carrier concentration and anomalous Hall effect due to a topological hybridization gap. 
The theoretical model exhibits two phases, a small Fermi surface magnet and a large Fermi surface heavy Fermi liquid; at the mean-field level the transition between them is first order. 
Our results provide concrete experimental predictions for ongoing experiments on MoTe$_2$/WSe$_2$ bilayer heterostructures and introduces a controlled route to observe a topological selective Mott transition.

\end{abstract}

\maketitle
\emph{Introduction.---}
Transition metal dichalcogenide (TMD) \mr devices created by stacking two TMD monolayers have recently emerged as a highly tunable platform to realize strongly correlated and topological states~\cite{Li_2021,Tao_202,Zhao_2022_Chern,TMD_Exper1,TMD_Exper2,TMD_Exper3,Wu_2018,Devakul_magic_2021,jiawei21,jiawei22,Jie21_staggered,AWietek21_stripeorder,Pan_2021,Devakul_2021,Xie_2022,Xie_preprint_2022,Xie_DasSarma_2022,Dong_2022,Margarita_LF_2206,PhysRevB.88.085433,kormanyos2015k}. 
This experimental flexibility has opened the door to control and observe phenomena that has been out of reach in conventional solid-state platforms such as a continuous Mott transition~\cite{TMD_Exper1,TMD_Exper2} in a single sample.
Forming heterobilayers with distinct chemical composition allows one to effectively tune the density and the interaction within an individual layer which can open the door to synthetically realize orbital selective Mott transitions~\cite{PhysRevLett.110.067003,Yi_2015,PhysRevB.94.115146,PhysRevLett.110.146402,Yi_2017}  to emulate several strongly correlated electron systems of interest.

\begin{figure}
    \centering
    \includegraphics[width=0.4\textwidth]{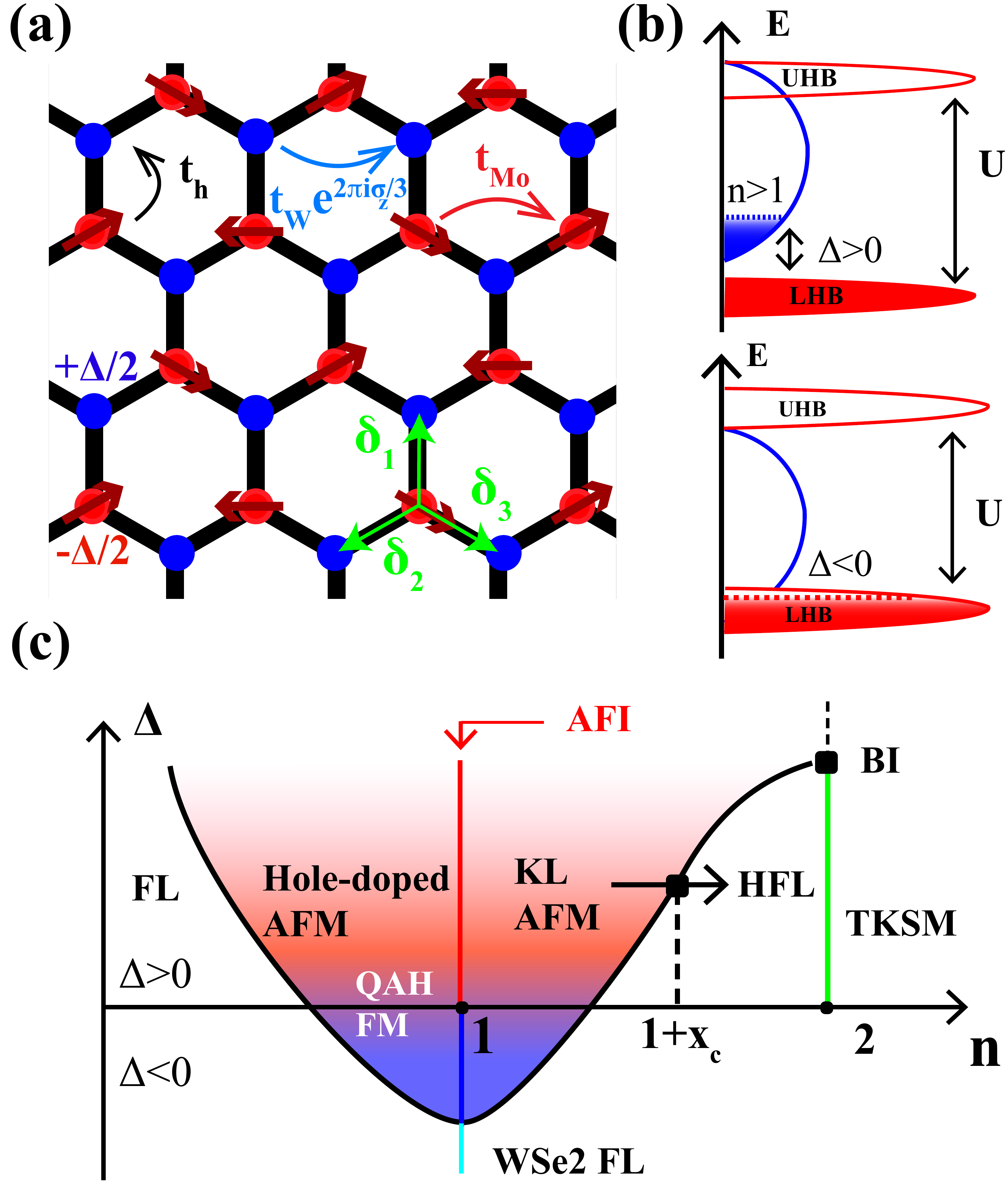}
    \vspace{-0.05cm}
    \caption{\textbf{Model and qualitative phase diagram}. (a) Representation of real-space structure of MoTe$_2$/WSe$_2$ bilayer showing the two triangular sublattices of the honeycomb moir\'e lattice. Red and blue represent the centers of the Wannier functions describing the relevant states in the bottom ($Mo$) and top ($W$) layers, respectively. The nearest neighbor vector $\bm\delta_i$, the $Mo-W$ energy difference $\Delta$ and the intralayer $t_W/t_{Mo}$ and interlayer $t_h$ hopping matrix elements are also indicated. (b) Schematic density of states for filling near $n=1$ representing the $Mo$ layer as a narrow-band strongly correlated system with lower (LHB) and upper (UHB) Hubbard bands separated by an energy gap $U$ while the $W$ layer is shown as a wide-band weakly correlated system. Upper panel: positive $\Delta$; showing filled $Mo$ LHB and  carriers added beyond $n=1$ in the $W$ layer. Lower panel: negative $\Delta$; mixed valent situation with potential for simultaneous occupancy of $Mo$- and $W$-layer states at $n=1$. (c) Qualitative phase diagram of the AB stacked MoTe2/WSe2 bilayer as a function of hole carrier density $n$ and displacement field $\Delta$ showing expected phases: at carrier density $n=1$ as $\Delta$ is decreased the insulating 120$^\circ$ antiferromagnet (AFM) gives way to a QAH insulator and then to a Fermi liquid (FL). For $n>1$ the the system can be described by a Kondo lattice (KL) model. At other carrier concentrations different metallic phases occur including moderate mass (FL) and heavy mass (HFL) Fermi liquids; a hole-doped single band antiferromagnet (AFM) and a Kondo lattice antiferromagnet (KLAFM) with ordered magnetism on the $Mo$ layer coupled to mobile carriers in the $W$ layer. The HFL is connected to a topological Kondo semimetal (TKSM) at $n=2$.}
    \label{fig:fig0}
\end{figure}

In this paper we focus on the AB-stacked MoTe$_2$/WSe$_2$ bilayer heterostructure of recent experimental~\cite{Li_2021} and theoretical~\cite{Pan_2021,Devakul_2021,Xie_2022,Xie_preprint_2022,Xie_DasSarma_2022,Dong_2022} interest and show it realizes a chiral Kondo lattice. In this system the lattice mismatch between the two materials leads to a hexagonal \mr lattice with a \mr lattice constant of $a_M\sim 5$nm. As shown in Fig.~\ref{fig:fig0}(a) the two sublattices of the hexagonal \mr lattice correspond to the MoTe$_2$ ($Mo$) and WSe$_2$ ($W$) layers, with the two sublattices connected by the interlayer hopping.

The combination of the strong spin-momentum locking of the monolayer materials and the AB stacking configuration reduces the magnitude of the interlayer tunnelling to a value much smaller than the intra-layer hopping of the WSe$_2$ \mr band ~\cite{Zhang_2021}. The small interlayer coupling means that the two layers can be discussed separately and then the effects of the interlayer coupling considered. The atomic physics of the monolayer materials determines that the MoTe$_2$ layer has a narrower \mr conduction band than does the WSe$_2$ layer, so that the MoTe$_2$ layer may be regarded as strongly correlated with an upper and lower Hubbard band while the WSe$_2$ layer has a wider bandwidth and a carrier concentration typically far from $n=1$ and may be regarded as weakly correlated~\cite{Zhang_2021,Devakul_2021}. The resulting density of states (DOS) is sketched in Fig.~\ref{fig:fig0}(b). The energy offset between the bands, defined here as $\Delta$, and the total chemical potential $\mu$ can be controlled {\em in situ} by appropriate gate voltages.  

Fig.~\ref{fig:fig0}(c) shows a qualitative phase diagram in the plane of band offset ($\Delta$) and total density ($n=1+x$). At large positive $\Delta$ the MoTe$_2$ band is well separated from the bottom of the WSe$_2$ band. The first carriers added to the system go into the MoTe$_2$ layer, forming a correlated metal which at $n=1$ becomes a triangular-lattice Mott insulator with $120^\circ$ antiferromagnetic (AFM) order. At carrier concentration $n=1$, decreasing $\Delta$ is predicted ~\cite{Zhang_2021,Devakul_2021,Xie_2022,Xie_preprint_2022,Xie_DasSarma_2022,Dong_2022} and observed~\cite{Li_2021,Tao_202} to cause a transition to a quantum anomalous Hall (QAH) state, followed by a transition to a conventional metallic state. At $\Delta>0$, carriers in excess of the Mott concentration at half-filling ($n=1$ per \mr unit cell) go into the WSe$_2$ band (if, as we assume, $U>\Delta$) and are coupled to the spins of the Mott insulator via an exchange coupling $J_K\sim t_h^2/\Delta$ derived perturbatively from the interlayer hybridization $t_h$, so that  the MoTe$_2$/WSe$_2$ system can be described by an effective Kondo lattice model on the \mr scale whose study is the central topic of this paper. 

Synthetic Kondo lattice models in \mr systems have been previously discussed in the context of the interplay between localized orbital and delocalized electrons in twisted bilayer~\cite{Song_hflTBG_2021} and trilayer graphene~\cite{Lado_2021}, and in relation to the orbital selective Mott transition in a two-band \mr TMD model~\cite{PhysRevResearch.3.043173}. More recently, a gate-tunable Kondo interaction in trilayer TMDs has been predicted to realize heavy fermion quantum criticality~\cite{Potter_2021}. Here, we focus on how the combination of strong spin-orbit coupling and the non-local structure of the interlayer hybridization substantially enriches the physics relative to the standard Kondo-lattice/orbitally selective Mott transition picture.

\emph{Derivation of the Kondo \mr lattice model.---}
The low-energy properties of the \mr system are described~\cite{Devakul_2021} by a Hubbard model on the honeycomb lattice shown in Fig.~\ref{fig:fig0}(a). The two sublattices of the honeycomb lattice give the centers of Wannier states formed from the $Mo$ (red) and $W$ (blue) sites respectively. Wannierization of band structure calculations~\cite{Devakul_2021} give same-sublattice hopping parameters $t_W\simeq9$meV, $t_{Mo}\simeq4.5$meV and an interlayer hopping $t_h\simeq2$meV shown as solid arrows in~\ref{fig:fig0}(a). The monolayer Ising spin-orbit coupling (SOC) implies that $t_W$ has a spin-dependent complex phase factor $\pm2\pi/3$ placing the W band minima at the Dirac points $\bm\kappa$ (spin up) and $\bm\kappa^\prime$ (spin down) respectively; for details see Ref.~\cite{Zhang_2021,Devakul_2021} or the SM~\cite{supplementary}. The local interaction $U$ which is taken to be the same on both layers for simplicity is believed to be large: $U/t_{Mo}\gg1$ \cite{Devakul_2021,Pan_2021} and for $\Delta>0$ gives rise at filling $n=1$ per \mr unit cell to a 120$^\circ$ AFM charge transfer insulator~\cite{Devakul_2021}. At nonzero electron doping $n=1+x$ $(x>0)$ the extra carriers go into the $W$ charger-transfer band (blue shaded region of DOS in Fig.~\ref{fig:fig0}(b)) and $Mo$ sites remain singly occupied. Due to the large bandwidth and the small doping $x$ we assume that the correlation effects in the $W$ band can be, at first approximation, ignored. The hybridization term $t_{h}$ induces an effective spin-exchange Kondo coupling~\cite{PhysRev.149.491,PhysRevB.37.9753,supplementary} between the dispersive electrons in the conduction band from the $W$ and the local moments $\bm S_\br$ from the $Mo$ layer. The resulting \mr Kondo-lattice (spin-fermion) model reads:
\begin{equation}
\label{Hsd}
\begin{split}
    \bar H =& \sum_{\bk\sigma}\xi_{\bk\sigma} c^\dagger_{\bk\sigma}c_{\bk\sigma} \\
    +& \sum_{\langle\br,\br'\rangle_{ \text{Mo}}}J_H\left(S^z_\br S^z_{\br'}+\gamma S^+_\br S^-_{\br'}+h.c.\right)+D (\bm{S}_{\br} \times \bm{S}_{\br'})_z \\
    +&\frac{1}{2N}\sum_{\br\in \text{Mo}}\sum_{\bk,\bp} e^{-i(\bk-\bp)\cdot\br}
    J_{\bk,\bp}
    \bm S_\br\cdot c^\dagger_{\bk \sigma} \,\bm \sigma_{\sigma\sigma'}\, c_{\bp\sigma'},
\end{split}
\end{equation}
where $\xi_{\bk\sigma}=-2t_W\sum^3_{j=1}\cos(\bk\cdot\bm a_i+2\pi s_\sigma/3)-\epsilon_F$ is the electron dispersion for the $W$ band,  $\bm a_1=\sqrt{3}a_{M}(1,0)$, $\bm a_{2,3}=\sqrt{3}a_{M}\left(-1/2,\pm\sqrt{3}/2\right)$ are the lattice vectors with $a_{M}=5$nm is the moir\'e-lattice constant, $s_{\uparrow,\downarrow}=\pm1$, and $\epsilon_F$ is the Fermi energy fixing the filling $x$ of electrons in the conduction band $\sum_{\br,\sigma} \langle c^\dagger_{\br\sigma} c_{\br\sigma}\rangle/N=x$. The exchange $J_H$ arises from a combination of the spin rotational invariant Heisenberg exchange in the $Mo$ layer and virtual excitations in the $W$ band mediated by $t_h$. The reduced spin rotational symmetry in the $W$ layer induces the XXZ anisotropy $\gamma\neq1/2$ and also Dzyaloshinskii–Moriya $D$ interactions~\cite{Devakul_magic_2021}. For simplicity we use the Heisenberg form $\gamma=1/2$ in Eq.~\eqn{Hsd} in our calculations and we treat $J_H$ as a phenomenological parameter.

In the last term of Eq.~\eqn{Hsd}, $J_{\bk,\bp}=J_K V^*_\bk V_\bp$ with $J_K=2t^2_{h}[1/\Delta+1/(U_\text{Mo}-\Delta)]$ is the Kondo exchange and the hopping $t_h$ between opposite sublattices gives the form factor $V_\bk=\sum_{j=1}^3 e^{i\bm\delta_j\cdot\bk}$ with $\bm\delta_j$ displayed as green arrows in Fig.~\ref{fig:fig0}(a). In proximity to the high-symmetry points $\bm\kappa$ and $\bm\kappa'$, centers of the spin up and down Fermi sea at low-doping, respectively, the form factor takes the $p$-wave chiral form $V_{\bm\kappa+\bk}\simeq-3a_M(k_x-ik_y)/2$ and due to time-reversal symmetry $V_{\bm\kappa'+\bk} = V^*_{\bm\kappa-\bk}$. The displacement field dependence ~\cite{Potter_2021} and chirality will be seen to have important physics consequences.

To quantify the Kondo coupling at the Fermi energy we introduce a Fermi surface (FS) averaged Kondo exchange $\bar{J}_K$ as:
\begin{equation}
    \bar J_K = \frac{J_K}{\rho_{\epsilon_F}}\frac{3\sqrt{3}a_M^2}{8\pi^2}\oint_{\rm FS} dk_t\frac{|V_\bk|^2}{|\nabla_\bk \epsilon_{\bk\sigma}|}\simeq \frac{xJ_K}{4\rho_{\epsilon_F} t_W},\label{defbarJK}
\end{equation}
where the line integral is over the conduction electron FS, $k_t$ is the component of $\bk$ along the tangent to the FS curve, $\rho_{\epsilon_F}$ is the DOS at the Fermi energy and in obtaining the explicit analytic form the $W$ band dispersion was expanded to leading (quadratic) order around its minimum.

\emph{Results and Methods.---}
In this section, we employ the mean-field theory of Abrikosov fermions~\cite{coleman1989kondo,Senthil_2004,Pixley_2014} to study the competition between the magnetic and the HFL phase. For this purpose we factorize the local magnetic spin into charge neutral spinons $\bm S_{\bm r} = \chi^{\dag}_{\bm r\alpha}\bm \sigma_{\alpha\beta}\chi_{\bm r\beta}/2$ subject to a constraint $\sum_{\sigma}\chi^{\dag}_{\bm r\sigma}\chi_{\bm r\sigma}=1$. In the following we adopt $\bm\sigma$ for the spin degrees of freedom. In our notation, $c_{\bm r}$ is a spin-doublet $(c_{\bm r\uparrow},c_{\bm r\downarrow})^T$ and same for $\chi_\br$. We treat the Kondo interaction with an unrestricted Hartree Fock ansatz that is equally split across the hybridization order parameter $\Phi$ and magnetic order $\bm M$ in the $Mo$ layer, which can induce a non-zero polarization in the $W$ layer $\bm m$. The HFL is captured by a non-vanishing amplitude of the hybridization $\Phi_{\bm r} \equiv -\sum_{j=1}^{3}\langle c^{\dag}_{\bm r+\bm\delta_j}\sigma_0\chi_{\bm r}\rangle/2$; the magnetic order is characterized by the variational parameters $\bm M_\br=(M^\parallel \cos\bQ\cdot\br,M^\parallel \sin\bQ\cdot\br,M^z)$ and $\bm m_\br=(m^\parallel \cos\bQ\cdot\br,m^\parallel \sin\bQ\cdot\br,m^z)$ with $\bQ=\bm\kappa-\bm\kappa'$, $M^\parallel=\sum_\bk[\langle\chi^\dagger_\bk\sigma_+\chi_{\bk+\bQ}\rangle+c.c.]/2N$, $M^z=\sum_\bk\langle\chi^\dagger_\bk\sigma_z\chi_\bk\rangle/2N$, $m^\parallel=\sum_\bk[V^*_\bk \langle c^\dagger_\bk\sigma_+ c_{\bk+\bQ}\rangle V_{\bk+\bQ}+c.c.]/2N$ and $m^z=\sum_\bk V^*_\bk \langle c^\dagger_\bk\sigma_zc_{\bk}\rangle V_\bk/2N$. To obtain the values of the variational parameters we minimize the mean-field free-energy as detailed in the SM~\cite{supplementary}.
\begin{figure}
    \centering
    \includegraphics[width=0.485\textwidth]{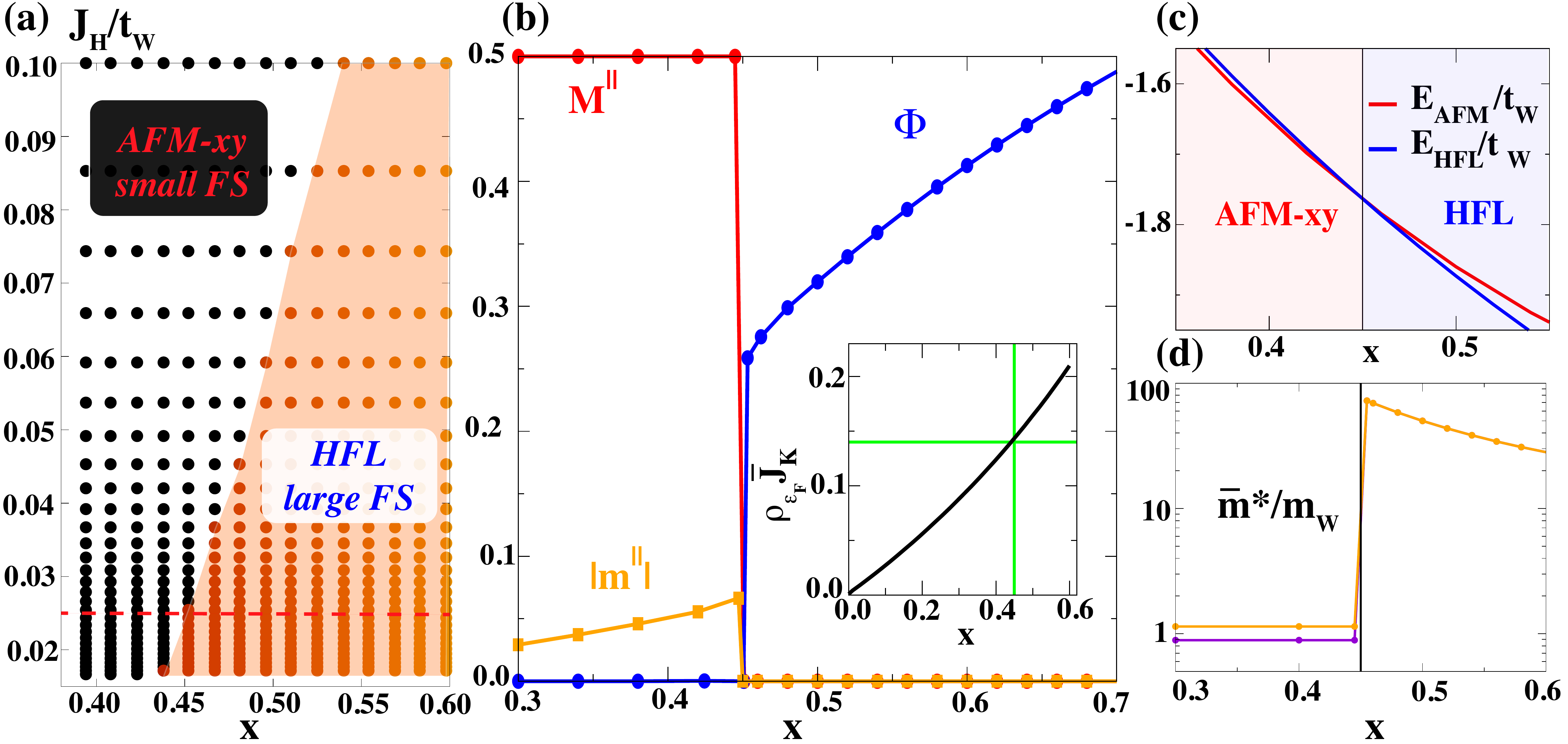}
    \vspace{-0.5cm}
    \caption{\textbf{Competition between HFL and magnetic order}. (a) Phase diagram in plane of WSe$_2$ filling $x$ vs $J_H/t_W$ calculated from solution of mean field equations. Blue does (white background) indicate regions of small Fermi surface metallic phase coexisting with 120$^\circ$ antiferromagnetic metallic order; red points (orange background) indicate regions of large Fermi surface heavy Fermi liquid phase. (b) Line cut at $J_H/t_W=0.025$ showing $x$-dependence of the mean-field order parameters: hybridization $\Phi$, local moment staggered magnetization $M^\parallel$ and itinerant electron staggered magnetization $|m^\parallel|$. $m^\parallel$ is opposite to $M^\parallel$;  we plot the absolute value of both quantities. The inset shows the effective Kondo exchange as a function of the doping $x$. A first order transition happens at the critical doping $x_c\simeq0.45$ corresponding to $\rho_{\epsilon_F}\bar J_K\simeq0.14$. (c) Energy of the two different phases as a function of $x$. (d) Evolution of the average quasiparticle mass as a function of $x$ in logarithmic scale. The data are obtained setting $J_{K}/t_W=1$ (bandwidth $9t_W$).}
    \label{fig:fig1}
\end{figure}

Fig.~\ref{fig:fig1}(a) shows the calculated mean-field phase diagram in the plane $x$ versus $J_H/t_W$. The evolution of the mean-field parameters along the line cut, red line in Fig.~\ref{fig:fig1}(a), is displayed in Fig.~\ref{fig:fig1}(b). In the small doping regime $x<x_c$ the mean-field minimum describes a magnetic solution with in-plane 120$^\circ$ AFM (AFM-xy). The antiferromagnetic sign of $J_K$ means that the conduction electron staggered magnetization $m^\parallel$ is directed oppositely to $\bm M$ shown as the orange line in Fig.~\ref{fig:fig1}(b). As doping $x$ increases,  $\rho_{\epsilon_F}\bar J_K$ grows as shown in the inset of Fig.~\ref{fig:fig1}(b), driving a transition of the general type discussed by Doniach \cite{Doniach_1977} to a non-magnetic Kondo lattice state in which the system becomes paramagnetic $\bm M=0$ and the localised moments hybridize with the conduction electrons giving rise to a large Fermi surface of heavy quasiparticles. 
 
The transition between the AFM-xy phase and the HFL is first order: both the AFM-xy order parameter and the hybridization change discontinuously and the computed energies cross [Fig.~\ref{fig:fig1}(c)]. Across the phase transition  the topology of the Fermi surface changes  from  electron like in the magnetic phase to hole like in the HFL phase. To gain insight on the nature of the transition, it is instructive to compare the low-doping behaviour $\rho_{\epsilon_F} \bar J_K\ll1$ of the two energy scales in the Kondo lattice model: the Kondo temperature $T_K$ and the magnetic energy $E_{AFM}$. Expanding to close to the bottom of the conduction band we have $T_K\simeq\epsilon_F\exp\left[-1/(\rho_{\epsilon_F} \bar J_K)\right]$ where for a quadratic dispersion the DOS is constant $\rho_{\epsilon_F}\simeq\rho_0=3\sqrt{3}a^2_M m_W/(4\pi h^2)$, $\epsilon_F\simeq x/(2\rho_0)$ and $m_W=\hbar^2/(9t_W a^2_M)$. On the other hand, in the magnetic state the Kondo coupling leads to a staggered polarization of conduction electrons which lowers the energy of the magnetic state: $E_{AFM}\simeq-3J_HM^2/2-\bar J^2_K\rho_{\epsilon_F}M^2/2$ ($M=1/2$). 
The linear dependence of $\bar J_K$ on  doping $x$ [Eq.~\eqn{defbarJK}] implies that the scaling of the magnetic to paramagnetic transition is different from the standard Doniach scaling~\cite{Doniach_1977}. 
Finally, Fig.~\ref{fig:fig1}(c) shows the evolution of the average over the FS of the quasiparticle mass $\bar m^*$ in logarithmic scale, where $m^*$ is defined as $m^*=\hbar |\bk_F|/|\bm v_F|$, $\bar m^*=\oint_{FS} dk_t m^*/(2\pi k_F)$ with $\bm v_F$ and $\bk_F$ Fermi velocity and momentum, respectively. The transition is signalled by a drastic change of the quasiparticle mass. Interestingly, we find a diminution of $\bar m^*$ increasing the doping $x$ and a splitting of the quasiparticle mass in the magnetic regime. 

\begin{figure}
    \centering
    \includegraphics[width=0.48\textwidth]{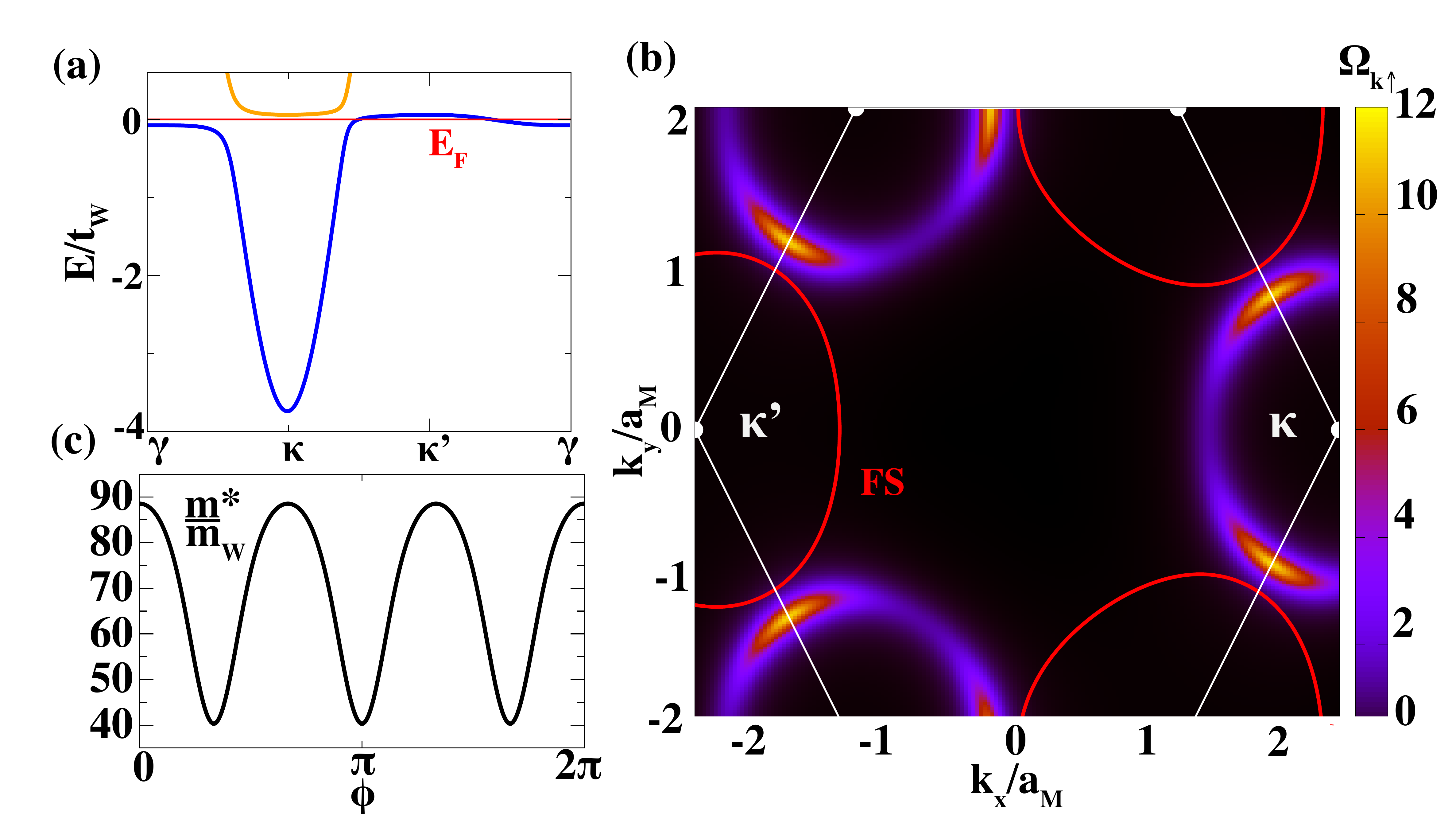}
    \vspace{-0.5cm}
    \caption{\textbf{Properties of the HFL}. (a) Band structure of the heavy Fermi liquid along high-symmetry directions where the chiral Kondo coupling opens a topological hybridization gap between the upper (orange) and lower (blue) bands. The energy is measured with respect to the chemical potential. (b) Berry curvature of the lower heavy Fermi liquid band, where the red circles denote the hole Fermi surface. The flux of $\Omega_{\bk\uparrow}$ in the entire Brillouin zone gives Chern number $C_\uparrow=1=-C_\downarrow$. (c) Quasiparticle effective mass $m^*/m_W$ as a function of the angle $\phi$ on the heavy electrons Fermi surface. The calculation has been performed at $x=0.46$ and $J_K/t_W=1$ (bandwidth $9t_W$).}
    \label{fig:fig2}
\end{figure}

\emph{Physical properties.---}
The HFL is characterized by the quasiparticle band structure in Fig.~\ref{fig:fig2}(a). In this regime the local moments in $Mo$ layer participate to the total volume~\cite{Oshikawa_2000} enclosed by the FS and give rise to a large hole-like FS which encircles the $\bm\kappa'(\bm\kappa)$ point for $\uparrow(\downarrow)$ electrons in the \mr Brillouin zone as shown by the solid red line in Fig.~\ref{fig:fig2}(b). We also show the variation of the quasiparticle mass around the Fermi surface in Fig.~\ref{fig:fig2}(c). In addition, we find quite unconventional properties that trace back to the form factor $V_\bk$ in the Kondo coupling $J_{\bk,\bp}$. The chiral nature of $V_\bk$ gives rise to a chiral hybridization order parameter whose amplitude is proportional to $\langle c^\dagger_{\bm\kappa+\bk \uparrow} \chi_{\bm\kappa+\bk \uparrow} \rangle\propto \Phi (k_x-ik_y)$ for spin up and $\langle c^\dagger_{\bm\kappa'+\bk \downarrow} \chi_{\bm\kappa'+\bk \downarrow} \rangle\propto -\Phi (k_x+ik_y)$ for spin down and results into the topological character of the hybridization gap. 
The color code in Fig.~\ref{fig:fig2}(b) shows the Berry curvature $\Omega_{\bk\uparrow}=-2\Im\braket{ \partial_{k_x}u_{\bk\uparrow}}{\partial_{k_y}u_{\bk\uparrow}}$ of the lower HFL band which is characterized by bright peaks at the position of the bare conduction electron FS. The opposite winding of spin $\uparrow$ and $\downarrow$ hybridization gap results in an opposite Berry curvature for the two spin $\Omega_{\bk\uparrow}=-\Omega_{-\bk\downarrow}$. 
The Berry curvature originates from the chiral interlayer hybridization between $\chi$ and $c$ fermions. Expanding the heavy Fermi liquid Hamiltonian around the electron pocket at $\bm\kappa$ for $\uparrow$ gives in the basis $\Psi=(\chi,c)$: 
\begin{equation}
    H_{\uparrow}(\bm\kappa+\bk)\simeq\left(\begin{matrix} \lambda & -\Delta_K (k_x-ik_y) \\  -\Delta^*_K (k_x+ik_y) & \hbar ^2k^2/(2m_W)-\mu \end{matrix}\right),
\end{equation}
with $\Delta_K=3J_Ka_M\Phi/2$ and due to time-reversal symmetry $H_{\downarrow}(\bm\kappa'+\bk)=H^*_{\uparrow}(\bm\kappa-\bk)$.
The HFL is adiabatically connected to a topological compensated Kondo semimetal at total filling $n=2$ with a non-quantized spin Hall effect~\cite{Kane_2005}. 
The topological gap found here is distinct from the topological Kondo gap in bulk heavy fermion systems~\cite{Dzero_2010,Dzero_2016} such as SmB$_6$~\cite{PhysRevB.88.180405,Kim_2013,Neupane_2013,Xu_2014}, which is induced by the the strong spin-orbit coupling and the opposite parity of $d$ and $f$ orbitals. In the system considered here the orbitals of conduction electrons and local moments have identical parity character and the topological hybridization gap originates from the non-local exchange involving nearest neighbour sites and the Ising SOC.

In the opposite regime, at $x<x_c$ the ground state has a 120$^\circ$ AFM order with a small Fermi surface. The spin-flip scattering processes mediated by the modulation $\bQ$ connecting the spin $\uparrow$ and $\downarrow$ FSs give rise to a SOC term in the conduction electron Hamiltonian. As detailed in the SM~\cite{supplementary} the expansion of the Hartree-Fock Hamiltonian close to the origin of the magnetic Brillouin zone gives:
\begin{equation}
\label{continuum_model}
\begin{split}
    \mathcal H_c(\bk)=\left(\frac{\hbar^2 k^2}{2m_W}-\epsilon_F\right)\sigma_0+\frac{9J_Ka^2_M }{8}\sum_{a}d_a(\bk)\sigma_a,
\end{split}
\end{equation}
where $d_z(\bk)=M^z(k^2_x+k^2_y)$, $d_x(\bk)=-M^\parallel(k^2_x-k^2_y)$ and $d_y(\bk)=2M^\parallel k_xk_y$. The SOC splits the conduction electron FS giving rise to two different Fermi momenta $k^\pm_{F}$. A nonzero in-plane $M^\parallel$ component in the Hamiltonian~\eqn{continuum_model} induces a $2\pi$-Berry phase winding around the origin of the magnetic Brillouin zone. In the presence of a nonzero Zeeman field we have a  Berry curvature whose momentum space distribution depends on the value of the parameters in the Hamiltonian~\eqn{continuum_model}. 

\begin{figure}
    \centering
    \includegraphics[width=0.475\textwidth]{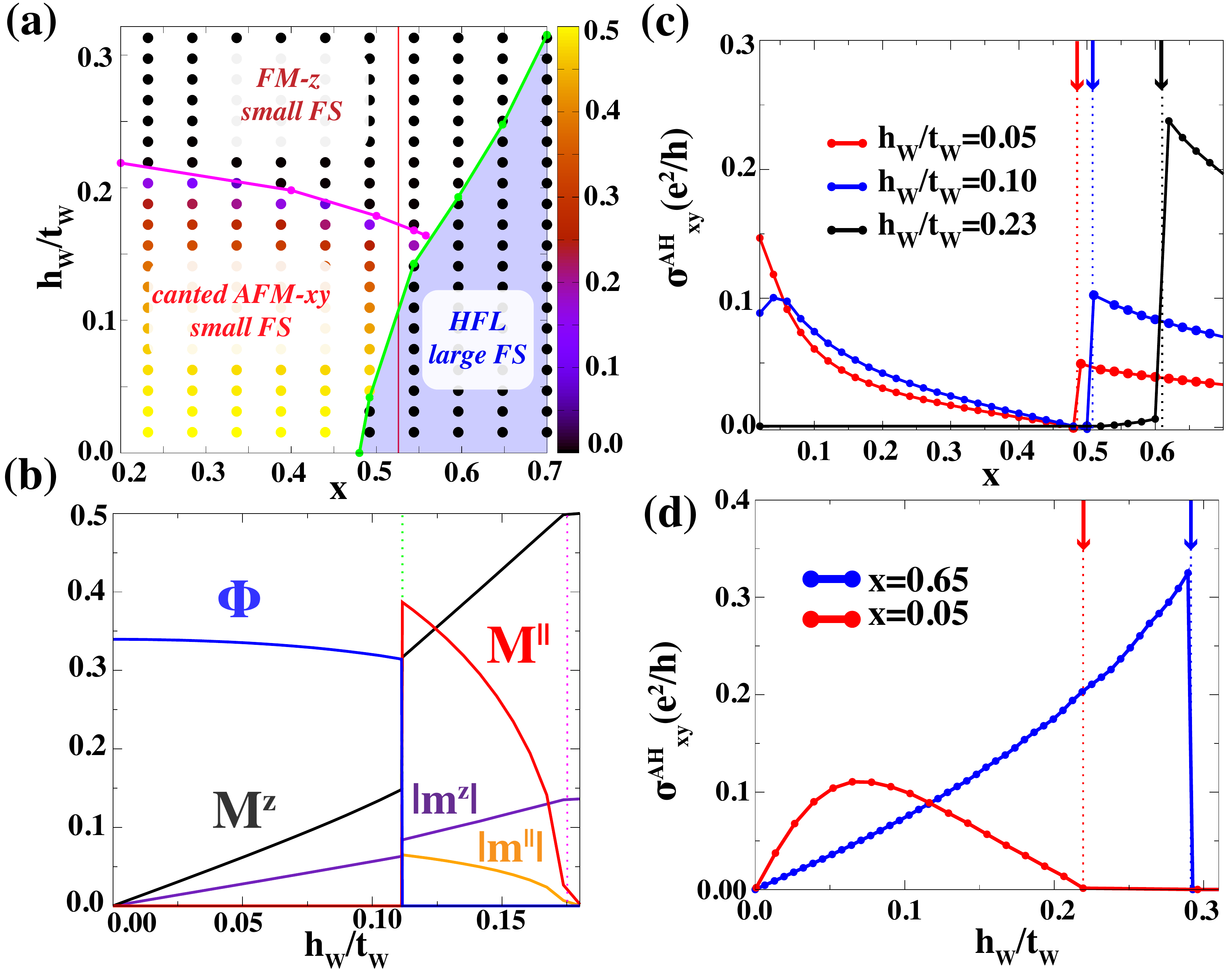}
    \vspace{-0.2cm}
    \caption{\textbf{Phase diagram and transport in magnetic field}.  (a) Phase diagram plotted as a function of the $W$-layer carrier concentration $x$ and the out-of-plane magnetic field $h_W$. The blue region denotes the HFL regime where the hybridization order parameters is finite. The color code before the green line shows the value of the in-plane antiferromagnetic order parameter $M^\parallel$. The solid green line shows the first order phase transition line. On the other hand, the magneta one shows the second order transition from the canted AFM-xy to the FM-z. (b) Evolution of the variational parameter along the line cut at $x=0.52$ as a function of $h_W/t_W$. (c) Intrinsic contribution to the anomalous Hall conductivity for three different values of the Zeeman field. The vertical dashed lines denote the critical filling where the first order transition takes place. (d) Intrinsic AHE for different values of the filling $x$ as a function of the Zeeman field. The vertical dashed line denotes the first order transition between the magnetic and the HFL phase, while the solid one the second order transition from a canted-AFM to the FM-z. Above this line the AHE vanishes. The calculation has been performed at $J_K/t_W=1$, $J_H/t_W=0.05$ (bandwidth $9t_W$).}
    \label{fig:fig3}
\end{figure}

Finally, we study the evolution of the mean-field solution in an external out-of-plane magnetic field. The spin-valley locking in the AB-stacked configuration implies that the magnetic field $B$ acts as a spin-valley Zeeman field~\cite{Zhao_2022_Chern} $H_{z}=-h_{Mo}\sum_{\br\in Mo} \chi^\dagger_{\br}\sigma^z\chi_{\br}/2+h_W\sum_{\br\in W}c^\dagger_\br \sigma_z c_\br/2$ where $h_\alpha=g_\alpha\mu_B B$ and $g_\alpha$ is the gyromagnetic ratio for the two different TMDs (here we use use $g_{Mo}=g_{W}=10$~\cite{Zhao_2022_Chern,PhysRevLett.126.067403}). 
The phase diagram in the plane $x$ vs $B$ is shown in Fig.~\ref{fig:fig3}(a). At low-doping a small magnetic field introduces an out-of-plane magnetization $M^z$ giving rise to a canted AFM-xy. 
Above the second order transition line (magenta line in Fig.~\ref{fig:fig3}(a)), that in the limit of vanishing doping takes the value $9 J_H/2$, the canted AFM-xy turns into a ferromagnetic solution along $z$ (FM-z). 
For larger fillings, blue region in Fig.~\ref{fig:fig3}(a) starting at $x_c\simeq0.488$ in the limit $h_W\to0$, the small-field solution is the HFL. The line cut Fig.~\ref{fig:fig3}(b), red line at $x=0.52$ in Fig.~\ref{fig:fig3}(a), shows the evolution of the variational parameters in magnetic field. In the HFL a small magnetic field induces a net magnetization $M^z$ in the localized orbital $\chi$. On the other hand, the Kondo hybridization is slightly affected by the external field. At a critical field displayed as a solid green line in Fig.~\ref{fig:fig3}(a) we find a first-order transition that is accompanied by an abrupt jump in the magnetization as well as by a discontinuous change in the conduction electron Fermi surface. The state above the first order line is a canted AFM-xy for $x<0.56$ and a FM-z for $x>0.56$.

\emph{Transport in magnetic field.---}
We use a Boltzmann equation approach to describe the transport of electrons in the different regimes of the phase diagram in Fig.~\ref{fig:fig3}(a), for details we refer to SM~\cite{supplementary}.  
The Hall conductivity has contribution from an Ohmic part $\sigma_{xy}^{\rm Ohm}$ which depends on the extrinsic impurity scattering rate, and also has an intrinsic geometric contribution~\cite{PhysRevB.43.193,Haldane_2004} $\sigma_{xy}^{\rm AH}$ determined by the Berry curvature. 
In the HFL regime the non-vanishing hybridization implies that the neutral $\chi$-spinon also contributes to the charge transport properties~\cite{Chowdhury_2018,Larkin_Ioffe_1989}. Approximating the HFL FS shown in Fig.~\ref{fig:fig2}(b) with a circular FS with effective mass $\bar{m}^*$ the Ohmic contribution takes the simple form $\sigma_{xx}=e^2\tau(1-x)/\bar{m}^*$ and $\sigma_{xy}^{\rm Ohm}=-e^3B\tau^2(1-x)/{\bar{m}^*}{}^2$. Conversely, in the magnetic regime only the concentration $x$ of conduction electrons contributes to the charge transport properties. Assuming a single momentum- and band-independent transport time we find $\sigma_{xx}=e^2\tau x/\left(\sum_\lambda m_\lambda/2\right)$ and $\sigma^{\rm Ohm}_{xy}=e^3\tau^2 B x/\left(m_W\sum_\lambda m_\lambda/2\right)$ with $m_\lambda=m_W/[1-\lambda J_K|M|/(4t_W)]\,(\lambda=\pm)$. As a result, increasing the magnetic field across the first order transition line [green line in Fig.~\ref{fig:fig3}(a)-(b)] we find a drastic jump in electrical conductivities, while in the HFL we have $1-x$ hole-like carriers in the AFM-xy phase a density of $x$ electrons. We observe that the normal Hall effect has been employed in heavy-Fermi liquid systems~\cite{Paschen_2004,Friedemann_pnas_2010,PhysRevB.82.035103} as a proxy for the Fermi surface change.  

Moreover, the AFM-xy and the HFL are characterized by different intrinsic anomalous Hall conductivity $\sigma^{\mathrm{AH}}_{xy}$. 
Figs.~\ref{fig:fig3}(c) shows the evolution of $\sigma^{\mathrm{AH}}_{xy}$ along different line cuts at constant Zeeman field as a function of doping $x$ in the phase diagram Fig.~\ref{fig:fig3}(a). In the AFM regime, $\sigma^{\mathrm{AH}}_{xy}$ originates from the $d_x(\bk)$ and $d_y(\bk)$ terms in Eq.~\eqn{continuum_model} induced by the in-plane component of the magnetic ordering which gives rise to the anomalous Hall effect (AHE) shown in Fig.~\ref{fig:fig3}(c). As we increase the doping the AHE decreases due to the smaller Berry curvature flux imbalance between the two Fermi surfaces with momenta $k^{\lambda}_F=\sqrt{2m_{\lambda}\epsilon_F/\hbar^2}$. 
The anomalous Hall effect is enhanced by interaction effects between conduction electrons which favours the out-of-plane canting of the AFM spin texture. 
Finally, above a critical density shown by the solid green line in the phase diagram Fig.~\ref{fig:fig3}(a) a first order transition from the canted AFM-xy to the HFL occurs.
The transition is signalled by a jump in the AHE effect as displayed in Fig.~\ref{fig:fig3}(c). A non-quantized jump of the AHE across the heavy Fermi liquid transition can also originate from a chiral spin liquid state~\cite{Ding_2015}. 
In the HFL regime, the anomalous Hall effect originates from the chiral hybridization gap with opposite sign for spin up and down electrons which results in a non-quantized spin Hall effect. See SM~\cite{supplementary} for details. A nonzero Zeeman field induces a flux imbalance between $\uparrow$ and $\downarrow$ which gives a finite AHE in the HFL as you can see in Fig.~\ref{fig:fig3}(c). 
In Fig.~\ref{fig:fig3}(d) we consider different concentration in the $W$-layer and we study the evolution as a function of $h_W$. We highlight that in the large field regime we enter in the FM-z regime where the AHE vanishes as displayed in the regions above the solid vertical lines in Fig.~\ref{fig:fig3}(c). 
The transition from the HFL to the FM-z is characterized transition by a drastic jump in the AHE effect which can reach values of the order of $\sim e^2/h$ depending on the flux imbalance between the two Fermi surfaces. On the other hand, in the small Fermi surface regime the AHE evolves smoothly vanishing when the local moments are aligned to the external field. 


\emph{Discussion.---}
We present a microscopic theory to explain the competition between AFM and the HFL in MoTe$_2$/WSe$_2$. 
A crucial finding is that the Kondo exchange is chiral, which gives rise to predictions that can be directly tested in experiments. Among them we emphasise the topological character of the hybridization gap in the heavy Fermi liquid regime and the Berry phase winding induced by spin-flip processes in the magnetic one. Both effects give rise to AHE which can be measured by transport experiments in magnetic field. We show how transport measurements can clearly distinguish the various phases.  

We find two different first order transitions:  either as a function of the doping $x$ at zero magnetic field, or by tuning  a magnetic field (metamagnetism).
The first order character of either transition is generically robust for a finite region of fluctuation corrections to mean field theory. However, while symmetry implies the magnetic field tuned transition is first order, there is no symmetry argument known to us that constrains the order of the doping tuned transition.
Our theory for the field tuned transition is in agreement with recent experimental observations~\cite{Zhao_2022} where at a critical field $B_c$ the heavy Fermi liquid undergoes a metamagnetic transition with a sharp jump in the carrier concentration from $x-1$ 
 to $x$ and an abrupt variation of the quasiparticle mass at a critical magnetic field $B_c$. We also find that at $B_c$ the transition between the magnetic state and the heavy Fermi liquid is signalled by a drastic jump of the AHE. 
 Additional experiments are needed to unveil the nature of the transition as a function of the doping at zero field.

We observe that the paramagnetic HFL solution becomes at filling $n=2$ a compensated topological Kondo semimetal with non-quantized quantum spin Hall effect. We also notice that in the magnetic regime the AHE can be further enhanced by spontaneous ferromagnetism induced by accounting the on-site interaction between conduction electrons. Finally, the effect of quantum fluctuations beyond the mean-field approach as well as the role of anisotropic effects in the Kondo exchange coupling are important questions for future research. 

These results highlight several concrete experimental predictions relevant for current and future experimental studies in TMDs bilayers. More, broadly this work provides a controlled route to realize a topological selective Mott transition.

\begin{acknowledgments}
\emph{Acknowledgements.---}
We thank Kin Fai Mak and Jie Shan for sharing with us experimental data prior to its publication and for insightful discussions. We have benefited from discussions with Elio K\"{o}nig, Piers Coleman, John Sous, Raquel Queiroz, Corentin Bertrand and Roberto Raimondi. This work was partially supported by the Air Force Office of Scientific Research under Grant No.~FA9550-20-1-0260 (J.C.) and Grant No.~FA9550-20-1-0136 (J.H.P.) and the Alfred P. Sloan Foundation through a Sloan Research Fellowship (J.H.P.). J.H.P. acknowledges the Aspen Center for Physics, where some of this work was completed, which is supported by National Science Foundation grant PHY-1607611. J.Z. and A.J.M. acknowledges support from the NSF MRSEC program through the Center for Precision-Assembled Quantum Materials (PAQM) - DMR-2011738. Flatiron Institute is a division of the Simons Foundation.

\emph{Author contributions.---}
DG, JW, JZ, JC JP and AM conceptualized the research. DG and JW developed the theoretical model and performed the numerical calculations. DG, JW, JZ, JC JP and AM analysed the results and edited the first draft.

\emph{Statement on competing interests.---}
All authors declare that they have no competing interests.

\emph{Data and materials availability.---}
All data needed to evaluate the conclusions in the paper are present in the paper and/or the Supplementary Materials.

\end{acknowledgments}

\bibliography{sample}

\begin{thebibliography}{57}%
\makeatletter
\providecommand \@ifxundefined [1]{%
 \@ifx{#1\undefined}
}%
\providecommand \@ifnum [1]{%
 \ifnum #1\expandafter \@firstoftwo
 \else \expandafter \@secondoftwo
 \fi
}%
\providecommand \@ifx [1]{%
 \ifx #1\expandafter \@firstoftwo
 \else \expandafter \@secondoftwo
 \fi
}%
\providecommand \natexlab [1]{#1}%
\providecommand \enquote  [1]{``#1''}%
\providecommand \bibnamefont  [1]{#1}%
\providecommand \bibfnamefont [1]{#1}%
\providecommand \citenamefont [1]{#1}%
\providecommand \href@noop [0]{\@secondoftwo}%
\providecommand \href [0]{\begingroup \@sanitize@url \@href}%
\providecommand \@href[1]{\@@startlink{#1}\@@href}%
\providecommand \@@href[1]{\endgroup#1\@@endlink}%
\providecommand \@sanitize@url [0]{\catcode `\\12\catcode `\$12\catcode
  `\&12\catcode `\#12\catcode `\^12\catcode `\_12\catcode `\%12\relax}%
\providecommand \@@startlink[1]{}%
\providecommand \@@endlink[0]{}%
\providecommand \url  [0]{\begingroup\@sanitize@url \@url }%
\providecommand \@url [1]{\endgroup\@href {#1}{\urlprefix }}%
\providecommand \urlprefix  [0]{URL }%
\providecommand \Eprint [0]{\href }%
\providecommand \doibase [0]{https://doi.org/}%
\providecommand \selectlanguage [0]{\@gobble}%
\providecommand \bibinfo  [0]{\@secondoftwo}%
\providecommand \bibfield  [0]{\@secondoftwo}%
\providecommand \translation [1]{[#1]}%
\providecommand \BibitemOpen [0]{}%
\providecommand \bibitemStop [0]{}%
\providecommand \bibitemNoStop [0]{.\EOS\space}%
\providecommand \EOS [0]{\spacefactor3000\relax}%
\providecommand \BibitemShut  [1]{\csname bibitem#1\endcsname}%
\let\auto@bib@innerbib\@empty
\bibitem [{\citenamefont {Li}\ \emph {et~al.}(2021{\natexlab{a}})\citenamefont
  {Li}, \citenamefont {Jiang}, \citenamefont {Shen}, \citenamefont {Zhang},
  \citenamefont {Li}, \citenamefont {Tao}, \citenamefont {Devakul},
  \citenamefont {Watanabe}, \citenamefont {Taniguchi}, \citenamefont {Fu},
  \citenamefont {Shan},\ and\ \citenamefont {Mak}}]{Li_2021}%
  \BibitemOpen
  \bibfield  {author} {\bibinfo {author} {\bibfnamefont {T.}~\bibnamefont
  {Li}}, \bibinfo {author} {\bibfnamefont {S.}~\bibnamefont {Jiang}}, \bibinfo
  {author} {\bibfnamefont {B.}~\bibnamefont {Shen}}, \bibinfo {author}
  {\bibfnamefont {Y.}~\bibnamefont {Zhang}}, \bibinfo {author} {\bibfnamefont
  {L.}~\bibnamefont {Li}}, \bibinfo {author} {\bibfnamefont {Z.}~\bibnamefont
  {Tao}}, \bibinfo {author} {\bibfnamefont {T.}~\bibnamefont {Devakul}},
  \bibinfo {author} {\bibfnamefont {K.}~\bibnamefont {Watanabe}}, \bibinfo
  {author} {\bibfnamefont {T.}~\bibnamefont {Taniguchi}}, \bibinfo {author}
  {\bibfnamefont {L.}~\bibnamefont {Fu}}, \bibinfo {author} {\bibfnamefont
  {J.}~\bibnamefont {Shan}},\ and\ \bibinfo {author} {\bibfnamefont {K.~F.}\
  \bibnamefont {Mak}},\ }\bibfield  {title} {\bibinfo {title} {Quantum
  anomalous hall effect from intertwined moir{\'{e}} bands},\ }\href
  {https://doi.org/10.1038/s41586-021-04171-1} {\bibfield  {journal} {\bibinfo
  {journal} {Nature}\ }\textbf {\bibinfo {volume} {600}},\ \bibinfo {pages}
  {641} (\bibinfo {year} {2021}{\natexlab{a}})}\BibitemShut {NoStop}%
\bibitem [{\citenamefont {Tao}\ \emph {et~al.}(2022)\citenamefont {Tao},
  \citenamefont {Shen}, \citenamefont {Jiang}, \citenamefont {Li},
  \citenamefont {Li}, \citenamefont {Ma}, \citenamefont {Zhao}, \citenamefont
  {Hu}, \citenamefont {Pistunova}, \citenamefont {Watanabe}, \citenamefont
  {Taniguchi}, \citenamefont {Heinz}, \citenamefont {Mak},\ and\ \citenamefont
  {Shan}}]{Tao_202}%
  \BibitemOpen
  \bibfield  {author} {\bibinfo {author} {\bibfnamefont {Z.}~\bibnamefont
  {Tao}}, \bibinfo {author} {\bibfnamefont {B.}~\bibnamefont {Shen}}, \bibinfo
  {author} {\bibfnamefont {S.}~\bibnamefont {Jiang}}, \bibinfo {author}
  {\bibfnamefont {T.}~\bibnamefont {Li}}, \bibinfo {author} {\bibfnamefont
  {L.}~\bibnamefont {Li}}, \bibinfo {author} {\bibfnamefont {L.}~\bibnamefont
  {Ma}}, \bibinfo {author} {\bibfnamefont {W.}~\bibnamefont {Zhao}}, \bibinfo
  {author} {\bibfnamefont {J.}~\bibnamefont {Hu}}, \bibinfo {author}
  {\bibfnamefont {K.}~\bibnamefont {Pistunova}}, \bibinfo {author}
  {\bibfnamefont {K.}~\bibnamefont {Watanabe}}, \bibinfo {author}
  {\bibfnamefont {T.}~\bibnamefont {Taniguchi}}, \bibinfo {author}
  {\bibfnamefont {T.~F.}\ \bibnamefont {Heinz}}, \bibinfo {author}
  {\bibfnamefont {K.~F.}\ \bibnamefont {Mak}},\ and\ \bibinfo {author}
  {\bibfnamefont {J.}~\bibnamefont {Shan}},\ }\href
  {https://doi.org/10.48550/ARXIV.2208.07452} {\bibinfo {title}
  {Valley-coherent quantum anomalous hall state in ab-stacked mote2/wse2
  bilayers}} (\bibinfo {year} {2022})\BibitemShut {NoStop}%
\bibitem [{\citenamefont {Zhao}\ \emph
  {et~al.}(2022{\natexlab{a}})\citenamefont {Zhao}, \citenamefont {Kang},
  \citenamefont {Li}, \citenamefont {Tschirhart}, \citenamefont {Redekop},
  \citenamefont {Watanabe}, \citenamefont {Taniguchi}, \citenamefont {Young},
  \citenamefont {Shan},\ and\ \citenamefont {Mak}}]{Zhao_2022_Chern}%
  \BibitemOpen
  \bibfield  {author} {\bibinfo {author} {\bibfnamefont {W.}~\bibnamefont
  {Zhao}}, \bibinfo {author} {\bibfnamefont {K.}~\bibnamefont {Kang}}, \bibinfo
  {author} {\bibfnamefont {L.}~\bibnamefont {Li}}, \bibinfo {author}
  {\bibfnamefont {C.}~\bibnamefont {Tschirhart}}, \bibinfo {author}
  {\bibfnamefont {E.}~\bibnamefont {Redekop}}, \bibinfo {author} {\bibfnamefont
  {K.}~\bibnamefont {Watanabe}}, \bibinfo {author} {\bibfnamefont
  {T.}~\bibnamefont {Taniguchi}}, \bibinfo {author} {\bibfnamefont
  {A.}~\bibnamefont {Young}}, \bibinfo {author} {\bibfnamefont
  {J.}~\bibnamefont {Shan}},\ and\ \bibinfo {author} {\bibfnamefont {K.~F.}\
  \bibnamefont {Mak}},\ }\href {https://doi.org/10.48550/ARXIV.2207.02312}
  {\bibinfo {title} {Realization of the haldane chern insulator in a moiré
  lattice}} (\bibinfo {year} {2022}{\natexlab{a}})\BibitemShut {NoStop}%
\bibitem [{\citenamefont {Li}\ \emph {et~al.}(2021{\natexlab{b}})\citenamefont
  {Li}, \citenamefont {Jiang}, \citenamefont {Li}, \citenamefont {Zhang},
  \citenamefont {Kang}, \citenamefont {Zhu}, \citenamefont {Watanabe},
  \citenamefont {Taniguchi}, \citenamefont {Chowdhury}, \citenamefont {Fu},
  \citenamefont {Shan},\ and\ \citenamefont {Mak}}]{TMD_Exper1}%
  \BibitemOpen
  \bibfield  {author} {\bibinfo {author} {\bibfnamefont {T.}~\bibnamefont
  {Li}}, \bibinfo {author} {\bibfnamefont {S.}~\bibnamefont {Jiang}}, \bibinfo
  {author} {\bibfnamefont {L.}~\bibnamefont {Li}}, \bibinfo {author}
  {\bibfnamefont {Y.}~\bibnamefont {Zhang}}, \bibinfo {author} {\bibfnamefont
  {K.}~\bibnamefont {Kang}}, \bibinfo {author} {\bibfnamefont {J.}~\bibnamefont
  {Zhu}}, \bibinfo {author} {\bibfnamefont {K.}~\bibnamefont {Watanabe}},
  \bibinfo {author} {\bibfnamefont {T.}~\bibnamefont {Taniguchi}}, \bibinfo
  {author} {\bibfnamefont {D.}~\bibnamefont {Chowdhury}}, \bibinfo {author}
  {\bibfnamefont {L.}~\bibnamefont {Fu}}, \bibinfo {author} {\bibfnamefont
  {J.}~\bibnamefont {Shan}},\ and\ \bibinfo {author} {\bibfnamefont {K.~F.}\
  \bibnamefont {Mak}},\ }\bibfield  {title} {\bibinfo {title} {Continuous mott
  transition in semiconductor moir{\'e}superlattices},\ }\href
  {https://doi.org/10.1038/s41586-021-03853-0} {\bibfield  {journal} {\bibinfo
  {journal} {Nature}\ }\textbf {\bibinfo {volume} {597}},\ \bibinfo {pages}
  {350} (\bibinfo {year} {2021}{\natexlab{b}})}\BibitemShut {NoStop}%
\bibitem [{\citenamefont {Ghiotto}\ \emph {et~al.}(2021)\citenamefont
  {Ghiotto}, \citenamefont {Shih}, \citenamefont {Pereira}, \citenamefont
  {Rhodes}, \citenamefont {Kim}, \citenamefont {Zang}, \citenamefont {Millis},
  \citenamefont {Watanabe}, \citenamefont {Taniguchi}, \citenamefont {Hone},
  \citenamefont {Wang}, \citenamefont {Dean},\ and\ \citenamefont
  {Pasupathy}}]{TMD_Exper2}%
  \BibitemOpen
  \bibfield  {author} {\bibinfo {author} {\bibfnamefont {A.}~\bibnamefont
  {Ghiotto}}, \bibinfo {author} {\bibfnamefont {E.-M.}\ \bibnamefont {Shih}},
  \bibinfo {author} {\bibfnamefont {G.~S. S.~G.}\ \bibnamefont {Pereira}},
  \bibinfo {author} {\bibfnamefont {D.~A.}\ \bibnamefont {Rhodes}}, \bibinfo
  {author} {\bibfnamefont {B.}~\bibnamefont {Kim}}, \bibinfo {author}
  {\bibfnamefont {J.}~\bibnamefont {Zang}}, \bibinfo {author} {\bibfnamefont
  {A.~J.}\ \bibnamefont {Millis}}, \bibinfo {author} {\bibfnamefont
  {K.}~\bibnamefont {Watanabe}}, \bibinfo {author} {\bibfnamefont
  {T.}~\bibnamefont {Taniguchi}}, \bibinfo {author} {\bibfnamefont {J.~C.}\
  \bibnamefont {Hone}}, \bibinfo {author} {\bibfnamefont {L.}~\bibnamefont
  {Wang}}, \bibinfo {author} {\bibfnamefont {C.~R.}\ \bibnamefont {Dean}},\
  and\ \bibinfo {author} {\bibfnamefont {A.~N.}\ \bibnamefont {Pasupathy}},\
  }\bibfield  {title} {\bibinfo {title} {Quantum criticality in twisted
  transition metal dichalcogenides},\ }\href
  {https://doi.org/10.1038/s41586-021-03815-6} {\bibfield  {journal} {\bibinfo
  {journal} {Nature}\ }\textbf {\bibinfo {volume} {597}},\ \bibinfo {pages}
  {345} (\bibinfo {year} {2021})}\BibitemShut {NoStop}%
\bibitem [{\citenamefont {Li}\ \emph {et~al.}(2021{\natexlab{c}})\citenamefont
  {Li}, \citenamefont {Li}, \citenamefont {Regan}, \citenamefont {Wang},
  \citenamefont {Zhao}, \citenamefont {Kahn}, \citenamefont {Yumigeta},
  \citenamefont {Blei}, \citenamefont {Taniguchi}, \citenamefont {Watanabe},
  \citenamefont {Tongay}, \citenamefont {Zettl}, \citenamefont {Crommie},\ and\
  \citenamefont {Wang}}]{TMD_Exper3}%
  \BibitemOpen
  \bibfield  {author} {\bibinfo {author} {\bibfnamefont {H.}~\bibnamefont
  {Li}}, \bibinfo {author} {\bibfnamefont {S.}~\bibnamefont {Li}}, \bibinfo
  {author} {\bibfnamefont {E.~C.}\ \bibnamefont {Regan}}, \bibinfo {author}
  {\bibfnamefont {D.}~\bibnamefont {Wang}}, \bibinfo {author} {\bibfnamefont
  {W.}~\bibnamefont {Zhao}}, \bibinfo {author} {\bibfnamefont {S.}~\bibnamefont
  {Kahn}}, \bibinfo {author} {\bibfnamefont {K.}~\bibnamefont {Yumigeta}},
  \bibinfo {author} {\bibfnamefont {M.}~\bibnamefont {Blei}}, \bibinfo {author}
  {\bibfnamefont {T.}~\bibnamefont {Taniguchi}}, \bibinfo {author}
  {\bibfnamefont {K.}~\bibnamefont {Watanabe}}, \bibinfo {author}
  {\bibfnamefont {S.}~\bibnamefont {Tongay}}, \bibinfo {author} {\bibfnamefont
  {A.}~\bibnamefont {Zettl}}, \bibinfo {author} {\bibfnamefont {M.~F.}\
  \bibnamefont {Crommie}},\ and\ \bibinfo {author} {\bibfnamefont
  {F.}~\bibnamefont {Wang}},\ }\bibfield  {title} {\bibinfo {title} {Imaging
  two-dimensional generalized wigner crystals},\ }\href
  {https://doi.org/10.1038/s41586-021-03874-9} {\bibfield  {journal} {\bibinfo
  {journal} {Nature}\ }\textbf {\bibinfo {volume} {597}},\ \bibinfo {pages}
  {650} (\bibinfo {year} {2021}{\natexlab{c}})}\BibitemShut {NoStop}%
\bibitem [{\citenamefont {Wu}\ \emph {et~al.}(2018)\citenamefont {Wu},
  \citenamefont {Lovorn}, \citenamefont {Tutuc},\ and\ \citenamefont
  {MacDonald}}]{Wu_2018}%
  \BibitemOpen
  \bibfield  {author} {\bibinfo {author} {\bibfnamefont {F.}~\bibnamefont
  {Wu}}, \bibinfo {author} {\bibfnamefont {T.}~\bibnamefont {Lovorn}}, \bibinfo
  {author} {\bibfnamefont {E.}~\bibnamefont {Tutuc}},\ and\ \bibinfo {author}
  {\bibfnamefont {A.~H.}\ \bibnamefont {MacDonald}},\ }\bibfield  {title}
  {\bibinfo {title} {Hubbard model physics in transition metal dichalcogenide
  moir\'e bands},\ }\href {https://doi.org/10.1103/PhysRevLett.121.026402}
  {\bibfield  {journal} {\bibinfo  {journal} {Phys. Rev. Lett.}\ }\textbf
  {\bibinfo {volume} {121}},\ \bibinfo {pages} {026402} (\bibinfo {year}
  {2018})}\BibitemShut {NoStop}%
\bibitem [{\citenamefont {Devakul}\ \emph {et~al.}(2021)\citenamefont
  {Devakul}, \citenamefont {Cr{\'{e}}pel}, \citenamefont {Zhang},\ and\
  \citenamefont {Fu}}]{Devakul_magic_2021}%
  \BibitemOpen
  \bibfield  {author} {\bibinfo {author} {\bibfnamefont {T.}~\bibnamefont
  {Devakul}}, \bibinfo {author} {\bibfnamefont {V.}~\bibnamefont
  {Cr{\'{e}}pel}}, \bibinfo {author} {\bibfnamefont {Y.}~\bibnamefont
  {Zhang}},\ and\ \bibinfo {author} {\bibfnamefont {L.}~\bibnamefont {Fu}},\
  }\bibfield  {title} {\bibinfo {title} {Magic in twisted transition metal
  dichalcogenide bilayers},\ }\bibfield  {journal} {\bibinfo  {journal} {Nature
  Communications}\ }\textbf {\bibinfo {volume} {12}},\ \href
  {https://doi.org/10.1038/s41467-021-27042-9} {10.1038/s41467-021-27042-9}
  (\bibinfo {year} {2021})\BibitemShut {NoStop}%
\bibitem [{\citenamefont {Zang}\ \emph {et~al.}(2021)\citenamefont {Zang},
  \citenamefont {Wang}, \citenamefont {Cano},\ and\ \citenamefont
  {Millis}}]{jiawei21}%
  \BibitemOpen
  \bibfield  {author} {\bibinfo {author} {\bibfnamefont {J.}~\bibnamefont
  {Zang}}, \bibinfo {author} {\bibfnamefont {J.}~\bibnamefont {Wang}}, \bibinfo
  {author} {\bibfnamefont {J.}~\bibnamefont {Cano}},\ and\ \bibinfo {author}
  {\bibfnamefont {A.~J.}\ \bibnamefont {Millis}},\ }\bibfield  {title}
  {\bibinfo {title} {Hartree-fock study of the moir\'e hubbard model for
  twisted bilayer transition metal dichalcogenides},\ }\href
  {https://doi.org/10.1103/PhysRevB.104.075150} {\bibfield  {journal} {\bibinfo
   {journal} {Phys. Rev. B}\ }\textbf {\bibinfo {volume} {104}},\ \bibinfo
  {pages} {075150} (\bibinfo {year} {2021})}\BibitemShut {NoStop}%
\bibitem [{\citenamefont {Zang}\ \emph {et~al.}(2022)\citenamefont {Zang},
  \citenamefont {Wang}, \citenamefont {Cano}, \citenamefont {Georges},\ and\
  \citenamefont {Millis}}]{jiawei22}%
  \BibitemOpen
  \bibfield  {author} {\bibinfo {author} {\bibfnamefont {J.}~\bibnamefont
  {Zang}}, \bibinfo {author} {\bibfnamefont {J.}~\bibnamefont {Wang}}, \bibinfo
  {author} {\bibfnamefont {J.}~\bibnamefont {Cano}}, \bibinfo {author}
  {\bibfnamefont {A.}~\bibnamefont {Georges}},\ and\ \bibinfo {author}
  {\bibfnamefont {A.~J.}\ \bibnamefont {Millis}},\ }\bibfield  {title}
  {\bibinfo {title} {Dynamical mean-field theory of moir\'e bilayer transition
  metal dichalcogenides: Phase diagram, resistivity, and quantum criticality},\
  }\href {https://doi.org/10.1103/PhysRevX.12.021064} {\bibfield  {journal}
  {\bibinfo  {journal} {Phys. Rev. X}\ }\textbf {\bibinfo {volume} {12}},\
  \bibinfo {pages} {021064} (\bibinfo {year} {2022})}\BibitemShut {NoStop}%
\bibitem [{\citenamefont {Wang}\ \emph {et~al.}(2023)\citenamefont {Wang},
  \citenamefont {Zang}, \citenamefont {Cano},\ and\ \citenamefont
  {Millis}}]{Jie21_staggered}%
  \BibitemOpen
  \bibfield  {author} {\bibinfo {author} {\bibfnamefont {J.}~\bibnamefont
  {Wang}}, \bibinfo {author} {\bibfnamefont {J.}~\bibnamefont {Zang}}, \bibinfo
  {author} {\bibfnamefont {J.}~\bibnamefont {Cano}},\ and\ \bibinfo {author}
  {\bibfnamefont {A.~J.}\ \bibnamefont {Millis}},\ }\bibfield  {title}
  {\bibinfo {title} {Staggered pseudo magnetic field in twisted transition
  metal dichalcogenides: Physical origin and experimental consequences},\
  }\href {https://doi.org/10.1103/PhysRevResearch.5.L012005} {\bibfield
  {journal} {\bibinfo  {journal} {Phys. Rev. Res.}\ }\textbf {\bibinfo {volume}
  {5}},\ \bibinfo {pages} {L012005} (\bibinfo {year} {2023})}\BibitemShut
  {NoStop}%
\bibitem [{\citenamefont {Wietek}\ \emph {et~al.}(2022)\citenamefont {Wietek},
  \citenamefont {Wang}, \citenamefont {Zang}, \citenamefont {Cano},
  \citenamefont {Georges},\ and\ \citenamefont
  {Millis}}]{AWietek21_stripeorder}%
  \BibitemOpen
  \bibfield  {author} {\bibinfo {author} {\bibfnamefont {A.}~\bibnamefont
  {Wietek}}, \bibinfo {author} {\bibfnamefont {J.}~\bibnamefont {Wang}},
  \bibinfo {author} {\bibfnamefont {J.}~\bibnamefont {Zang}}, \bibinfo {author}
  {\bibfnamefont {J.}~\bibnamefont {Cano}}, \bibinfo {author} {\bibfnamefont
  {A.}~\bibnamefont {Georges}},\ and\ \bibinfo {author} {\bibfnamefont
  {A.}~\bibnamefont {Millis}},\ }\bibfield  {title} {\bibinfo {title} {Tunable
  stripe order and weak superconductivity in the moir\'e hubbard model},\
  }\href {https://doi.org/10.1103/PhysRevResearch.4.043048} {\bibfield
  {journal} {\bibinfo  {journal} {Phys. Rev. Res.}\ }\textbf {\bibinfo {volume}
  {4}},\ \bibinfo {pages} {043048} (\bibinfo {year} {2022})}\BibitemShut
  {NoStop}%
\bibitem [{\citenamefont {Pan}\ \emph {et~al.}(2021)\citenamefont {Pan},
  \citenamefont {Xie}, \citenamefont {Wu},\ and\ \citenamefont
  {Sarma}}]{Pan_2021}%
  \BibitemOpen
  \bibfield  {author} {\bibinfo {author} {\bibfnamefont {H.}~\bibnamefont
  {Pan}}, \bibinfo {author} {\bibfnamefont {M.}~\bibnamefont {Xie}}, \bibinfo
  {author} {\bibfnamefont {F.}~\bibnamefont {Wu}},\ and\ \bibinfo {author}
  {\bibfnamefont {S.~D.}\ \bibnamefont {Sarma}},\ }\href
  {https://doi.org/10.48550/ARXIV.2111.01152} {\bibinfo {title} {Topological
  phases in ab-stacked mote$_2$/wse$_2$: $\mathbb{Z}_2$ topological insulators,
  chern insulators, and topological charge density waves}} (\bibinfo {year}
  {2021})\BibitemShut {NoStop}%
\bibitem [{\citenamefont {Devakul}\ and\ \citenamefont
  {Fu}(2021)}]{Devakul_2021}%
  \BibitemOpen
  \bibfield  {author} {\bibinfo {author} {\bibfnamefont {T.}~\bibnamefont
  {Devakul}}\ and\ \bibinfo {author} {\bibfnamefont {L.}~\bibnamefont {Fu}},\
  }\href {https://doi.org/10.48550/ARXIV.2109.13909} {\bibinfo {title} {Quantum
  anomalous hall effect from inverted charge transfer gap}} (\bibinfo {year}
  {2021})\BibitemShut {NoStop}%
\bibitem [{\citenamefont {Xie}\ \emph {et~al.}(2022{\natexlab{a}})\citenamefont
  {Xie}, \citenamefont {Zhang}, \citenamefont {Hu}, \citenamefont {Mak},\ and\
  \citenamefont {Law}}]{Xie_2022}%
  \BibitemOpen
  \bibfield  {author} {\bibinfo {author} {\bibfnamefont {Y.-M.}\ \bibnamefont
  {Xie}}, \bibinfo {author} {\bibfnamefont {C.-P.}\ \bibnamefont {Zhang}},
  \bibinfo {author} {\bibfnamefont {J.-X.}\ \bibnamefont {Hu}}, \bibinfo
  {author} {\bibfnamefont {K.~F.}\ \bibnamefont {Mak}},\ and\ \bibinfo {author}
  {\bibfnamefont {K.~T.}\ \bibnamefont {Law}},\ }\bibfield  {title} {\bibinfo
  {title} {Valley-polarized quantum anomalous hall state in moir\'e
  ${\mathrm{mote}}_{2}/{\mathrm{wse}}_{2}$ heterobilayers},\ }\href
  {https://doi.org/10.1103/PhysRevLett.128.026402} {\bibfield  {journal}
  {\bibinfo  {journal} {Phys. Rev. Lett.}\ }\textbf {\bibinfo {volume} {128}},\
  \bibinfo {pages} {026402} (\bibinfo {year} {2022}{\natexlab{a}})}\BibitemShut
  {NoStop}%
\bibitem [{\citenamefont {Xie}\ \emph {et~al.}(2022{\natexlab{b}})\citenamefont
  {Xie}, \citenamefont {Zhang},\ and\ \citenamefont {Law}}]{Xie_preprint_2022}%
  \BibitemOpen
  \bibfield  {author} {\bibinfo {author} {\bibfnamefont {Y.-M.}\ \bibnamefont
  {Xie}}, \bibinfo {author} {\bibfnamefont {C.-P.}\ \bibnamefont {Zhang}},\
  and\ \bibinfo {author} {\bibfnamefont {K.~T.}\ \bibnamefont {Law}},\ }\href
  {https://doi.org/10.48550/ARXIV.2206.11666} {\bibinfo {title} {Topological
  $p_x+ip_y$ inter-valley coherent state in moiré mote$_2$/wse$_2$
  heterobilayers}} (\bibinfo {year} {2022}{\natexlab{b}})\BibitemShut {NoStop}%
\bibitem [{\citenamefont {Xie}\ \emph {et~al.}(2022{\natexlab{c}})\citenamefont
  {Xie}, \citenamefont {Pan}, \citenamefont {Wu},\ and\ \citenamefont
  {Sarma}}]{Xie_DasSarma_2022}%
  \BibitemOpen
  \bibfield  {author} {\bibinfo {author} {\bibfnamefont {M.}~\bibnamefont
  {Xie}}, \bibinfo {author} {\bibfnamefont {H.}~\bibnamefont {Pan}}, \bibinfo
  {author} {\bibfnamefont {F.}~\bibnamefont {Wu}},\ and\ \bibinfo {author}
  {\bibfnamefont {S.~D.}\ \bibnamefont {Sarma}},\ }\href
  {https://doi.org/10.48550/ARXIV.2206.12427} {\bibinfo {title} {Nematic
  excitonic insulator in transition metal dichalcogenide moiré
  heterobilayers}} (\bibinfo {year} {2022}{\natexlab{c}})\BibitemShut {NoStop}%
\bibitem [{\citenamefont {Dong}\ and\ \citenamefont {Zhang}(2022)}]{Dong_2022}%
  \BibitemOpen
  \bibfield  {author} {\bibinfo {author} {\bibfnamefont {Z.}~\bibnamefont
  {Dong}}\ and\ \bibinfo {author} {\bibfnamefont {Y.-H.}\ \bibnamefont
  {Zhang}},\ }\href {https://doi.org/10.48550/ARXIV.2206.13567} {\bibinfo
  {title} {Excitonic chern insulator and kinetic ferromagnetism in
  mote$_2$/wse$_2$ moiré bilayer}} (\bibinfo {year} {2022})\BibitemShut
  {NoStop}%
\bibitem [{\citenamefont {{Davydova}}\ \emph {et~al.}(2022)\citenamefont
  {{Davydova}}, \citenamefont {{Zhang}},\ and\ \citenamefont
  {{Fu}}}]{Margarita_LF_2206}%
  \BibitemOpen
  \bibfield  {author} {\bibinfo {author} {\bibfnamefont {M.}~\bibnamefont
  {{Davydova}}}, \bibinfo {author} {\bibfnamefont {Y.}~\bibnamefont
  {{Zhang}}},\ and\ \bibinfo {author} {\bibfnamefont {L.}~\bibnamefont
  {{Fu}}},\ }\bibfield  {title} {\bibinfo {title} {{Itinerant spin polaron and
  metallic ferromagnetism in semiconductor moir{\'e} superlattices}},\
  }\href@noop {} {\bibfield  {journal} {\bibinfo  {journal} {arXiv e-prints}\
  ,\ \bibinfo {eid} {arXiv:2206.01221}} (\bibinfo {year} {2022})},\ \Eprint
  {https://arxiv.org/abs/2206.01221} {arXiv:2206.01221 [cond-mat.str-el]}
  \BibitemShut {NoStop}%
\bibitem [{\citenamefont {Liu}\ \emph {et~al.}(2013)\citenamefont {Liu},
  \citenamefont {Shan}, \citenamefont {Yao}, \citenamefont {Yao},\ and\
  \citenamefont {Xiao}}]{PhysRevB.88.085433}%
  \BibitemOpen
  \bibfield  {author} {\bibinfo {author} {\bibfnamefont {G.-B.}\ \bibnamefont
  {Liu}}, \bibinfo {author} {\bibfnamefont {W.-Y.}\ \bibnamefont {Shan}},
  \bibinfo {author} {\bibfnamefont {Y.}~\bibnamefont {Yao}}, \bibinfo {author}
  {\bibfnamefont {W.}~\bibnamefont {Yao}},\ and\ \bibinfo {author}
  {\bibfnamefont {D.}~\bibnamefont {Xiao}},\ }\bibfield  {title} {\bibinfo
  {title} {Three-band tight-binding model for monolayers of group-vib
  transition metal dichalcogenides},\ }\href
  {https://doi.org/10.1103/PhysRevB.88.085433} {\bibfield  {journal} {\bibinfo
  {journal} {Phys. Rev. B}\ }\textbf {\bibinfo {volume} {88}},\ \bibinfo
  {pages} {085433} (\bibinfo {year} {2013})}\BibitemShut {NoStop}%
\bibitem [{\citenamefont {Korm{\'a}nyos}\ \emph {et~al.}(2015)\citenamefont
  {Korm{\'a}nyos}, \citenamefont {Burkard}, \citenamefont {Gmitra},
  \citenamefont {Fabian}, \citenamefont {Z{\'o}lyomi}, \citenamefont
  {Drummond},\ and\ \citenamefont {Fal’ko}}]{kormanyos2015k}%
  \BibitemOpen
  \bibfield  {author} {\bibinfo {author} {\bibfnamefont {A.}~\bibnamefont
  {Korm{\'a}nyos}}, \bibinfo {author} {\bibfnamefont {G.}~\bibnamefont
  {Burkard}}, \bibinfo {author} {\bibfnamefont {M.}~\bibnamefont {Gmitra}},
  \bibinfo {author} {\bibfnamefont {J.}~\bibnamefont {Fabian}}, \bibinfo
  {author} {\bibfnamefont {V.}~\bibnamefont {Z{\'o}lyomi}}, \bibinfo {author}
  {\bibfnamefont {N.~D.}\ \bibnamefont {Drummond}},\ and\ \bibinfo {author}
  {\bibfnamefont {V.}~\bibnamefont {Fal’ko}},\ }\bibfield  {title} {\bibinfo
  {title} {k{\textperiodcentered} p theory for two-dimensional transition metal
  dichalcogenide semiconductors},\ }\href@noop {} {\bibfield  {journal}
  {\bibinfo  {journal} {2D Materials}\ }\textbf {\bibinfo {volume} {2}},\
  \bibinfo {pages} {022001} (\bibinfo {year} {2015})}\BibitemShut {NoStop}%
\bibitem [{\citenamefont {Yi}\ \emph {et~al.}(2013)\citenamefont {Yi},
  \citenamefont {Lu}, \citenamefont {Yu}, \citenamefont {Riggs}, \citenamefont
  {Chu}, \citenamefont {Lv}, \citenamefont {Liu}, \citenamefont {Lu},
  \citenamefont {Cui}, \citenamefont {Hashimoto}, \citenamefont {Mo},
  \citenamefont {Hussain}, \citenamefont {Chu}, \citenamefont {Fisher},
  \citenamefont {Si},\ and\ \citenamefont {Shen}}]{PhysRevLett.110.067003}%
  \BibitemOpen
  \bibfield  {author} {\bibinfo {author} {\bibfnamefont {M.}~\bibnamefont
  {Yi}}, \bibinfo {author} {\bibfnamefont {D.~H.}\ \bibnamefont {Lu}}, \bibinfo
  {author} {\bibfnamefont {R.}~\bibnamefont {Yu}}, \bibinfo {author}
  {\bibfnamefont {S.~C.}\ \bibnamefont {Riggs}}, \bibinfo {author}
  {\bibfnamefont {J.-H.}\ \bibnamefont {Chu}}, \bibinfo {author} {\bibfnamefont
  {B.}~\bibnamefont {Lv}}, \bibinfo {author} {\bibfnamefont {Z.~K.}\
  \bibnamefont {Liu}}, \bibinfo {author} {\bibfnamefont {M.}~\bibnamefont
  {Lu}}, \bibinfo {author} {\bibfnamefont {Y.-T.}\ \bibnamefont {Cui}},
  \bibinfo {author} {\bibfnamefont {M.}~\bibnamefont {Hashimoto}}, \bibinfo
  {author} {\bibfnamefont {S.-K.}\ \bibnamefont {Mo}}, \bibinfo {author}
  {\bibfnamefont {Z.}~\bibnamefont {Hussain}}, \bibinfo {author} {\bibfnamefont
  {C.~W.}\ \bibnamefont {Chu}}, \bibinfo {author} {\bibfnamefont {I.~R.}\
  \bibnamefont {Fisher}}, \bibinfo {author} {\bibfnamefont {Q.}~\bibnamefont
  {Si}},\ and\ \bibinfo {author} {\bibfnamefont {Z.-X.}\ \bibnamefont {Shen}},\
  }\bibfield  {title} {\bibinfo {title} {Observation of temperature-induced
  crossover to an orbital-selective mott phase in
  ${\mathrm{a}}_{x}{\mathrm{fe}}_{2\mathrm{\text{\ensuremath{-}}}y}{\mathrm{se}}_{2}$
  ($a\mathbf{=}\mathrm{K}$, rb) superconductors},\ }\href
  {https://doi.org/10.1103/PhysRevLett.110.067003} {\bibfield  {journal}
  {\bibinfo  {journal} {Phys. Rev. Lett.}\ }\textbf {\bibinfo {volume} {110}},\
  \bibinfo {pages} {067003} (\bibinfo {year} {2013})}\BibitemShut {NoStop}%
\bibitem [{\citenamefont {Yi}\ \emph {et~al.}(2015)\citenamefont {Yi},
  \citenamefont {Liu}, \citenamefont {Zhang}, \citenamefont {Yu}, \citenamefont
  {Zhu}, \citenamefont {Lee}, \citenamefont {Moore}, \citenamefont {Schmitt},
  \citenamefont {Li}, \citenamefont {Riggs}, \citenamefont {Chu}, \citenamefont
  {Lv}, \citenamefont {Hu}, \citenamefont {Hashimoto}, \citenamefont {Mo},
  \citenamefont {Hussain}, \citenamefont {Mao}, \citenamefont {Chu},
  \citenamefont {Fisher}, \citenamefont {Si}, \citenamefont {Shen},\ and\
  \citenamefont {Lu}}]{Yi_2015}%
  \BibitemOpen
  \bibfield  {author} {\bibinfo {author} {\bibfnamefont {M.}~\bibnamefont
  {Yi}}, \bibinfo {author} {\bibfnamefont {Z.-K.}\ \bibnamefont {Liu}},
  \bibinfo {author} {\bibfnamefont {Y.}~\bibnamefont {Zhang}}, \bibinfo
  {author} {\bibfnamefont {R.}~\bibnamefont {Yu}}, \bibinfo {author}
  {\bibfnamefont {J.-X.}\ \bibnamefont {Zhu}}, \bibinfo {author} {\bibfnamefont
  {J.}~\bibnamefont {Lee}}, \bibinfo {author} {\bibfnamefont {R.}~\bibnamefont
  {Moore}}, \bibinfo {author} {\bibfnamefont {F.}~\bibnamefont {Schmitt}},
  \bibinfo {author} {\bibfnamefont {W.}~\bibnamefont {Li}}, \bibinfo {author}
  {\bibfnamefont {S.}~\bibnamefont {Riggs}}, \bibinfo {author} {\bibfnamefont
  {J.-H.}\ \bibnamefont {Chu}}, \bibinfo {author} {\bibfnamefont
  {B.}~\bibnamefont {Lv}}, \bibinfo {author} {\bibfnamefont {J.}~\bibnamefont
  {Hu}}, \bibinfo {author} {\bibfnamefont {M.}~\bibnamefont {Hashimoto}},
  \bibinfo {author} {\bibfnamefont {S.-K.}\ \bibnamefont {Mo}}, \bibinfo
  {author} {\bibfnamefont {Z.}~\bibnamefont {Hussain}}, \bibinfo {author}
  {\bibfnamefont {Z.}~\bibnamefont {Mao}}, \bibinfo {author} {\bibfnamefont
  {C.}~\bibnamefont {Chu}}, \bibinfo {author} {\bibfnamefont {I.}~\bibnamefont
  {Fisher}}, \bibinfo {author} {\bibfnamefont {Q.}~\bibnamefont {Si}}, \bibinfo
  {author} {\bibfnamefont {Z.-X.}\ \bibnamefont {Shen}},\ and\ \bibinfo
  {author} {\bibfnamefont {D.}~\bibnamefont {Lu}},\ }\bibfield  {title}
  {\bibinfo {title} {Observation of universal strong orbital-dependent
  correlation effects in iron chalcogenides},\ }\bibfield  {journal} {\bibinfo
  {journal} {Nature Communications}\ }\textbf {\bibinfo {volume} {6}},\ \href
  {https://doi.org/10.1038/ncomms8777} {10.1038/ncomms8777} (\bibinfo {year}
  {2015})\BibitemShut {NoStop}%
\bibitem [{\citenamefont {Pu}\ \emph {et~al.}(2016)\citenamefont {Pu},
  \citenamefont {Huang}, \citenamefont {Xu}, \citenamefont {Xu}, \citenamefont
  {Song}, \citenamefont {Wen}, \citenamefont {Peng},\ and\ \citenamefont
  {Feng}}]{PhysRevB.94.115146}%
  \BibitemOpen
  \bibfield  {author} {\bibinfo {author} {\bibfnamefont {Y.~J.}\ \bibnamefont
  {Pu}}, \bibinfo {author} {\bibfnamefont {Z.~C.}\ \bibnamefont {Huang}},
  \bibinfo {author} {\bibfnamefont {H.~C.}\ \bibnamefont {Xu}}, \bibinfo
  {author} {\bibfnamefont {D.~F.}\ \bibnamefont {Xu}}, \bibinfo {author}
  {\bibfnamefont {Q.}~\bibnamefont {Song}}, \bibinfo {author} {\bibfnamefont
  {C.~H.~P.}\ \bibnamefont {Wen}}, \bibinfo {author} {\bibfnamefont
  {R.}~\bibnamefont {Peng}},\ and\ \bibinfo {author} {\bibfnamefont {D.~L.}\
  \bibnamefont {Feng}},\ }\bibfield  {title} {\bibinfo {title}
  {Temperature-induced orbital selective localization and coherent-incoherent
  crossover in single-layer
  $\mathrm{FeSe}/\mathrm{Nb}:{\mathrm{batio}}_{3}/{\mathrm{ktao}}_{3}$},\
  }\href {https://doi.org/10.1103/PhysRevB.94.115146} {\bibfield  {journal}
  {\bibinfo  {journal} {Phys. Rev. B}\ }\textbf {\bibinfo {volume} {94}},\
  \bibinfo {pages} {115146} (\bibinfo {year} {2016})}\BibitemShut {NoStop}%
\bibitem [{\citenamefont {Yu}\ and\ \citenamefont
  {Si}(2013)}]{PhysRevLett.110.146402}%
  \BibitemOpen
  \bibfield  {author} {\bibinfo {author} {\bibfnamefont {R.}~\bibnamefont
  {Yu}}\ and\ \bibinfo {author} {\bibfnamefont {Q.}~\bibnamefont {Si}},\
  }\bibfield  {title} {\bibinfo {title} {Orbital-selective mott phase in
  multiorbital models for alkaline iron selenides
  ${\mathbf{k}}_{1\ensuremath{-}x}{\mathrm{fe}}_{2\ensuremath{-}y}{\mathrm{se}}_{2}$},\
  }\href {https://doi.org/10.1103/PhysRevLett.110.146402} {\bibfield  {journal}
  {\bibinfo  {journal} {Phys. Rev. Lett.}\ }\textbf {\bibinfo {volume} {110}},\
  \bibinfo {pages} {146402} (\bibinfo {year} {2013})}\BibitemShut {NoStop}%
\bibitem [{\citenamefont {{Yi}}\ \emph {et~al.}(2017)\citenamefont {{Yi}},
  \citenamefont {{Zhang}}, \citenamefont {{Shen}},\ and\ \citenamefont
  {{Lu}}}]{Yi_2017}%
  \BibitemOpen
  \bibfield  {author} {\bibinfo {author} {\bibfnamefont {M.}~\bibnamefont
  {{Yi}}}, \bibinfo {author} {\bibfnamefont {Y.}~\bibnamefont {{Zhang}}},
  \bibinfo {author} {\bibfnamefont {Z.-X.}\ \bibnamefont {{Shen}}},\ and\
  \bibinfo {author} {\bibfnamefont {D.}~\bibnamefont {{Lu}}},\ }\bibfield
  {title} {\bibinfo {title} {{Role of the orbital degree of freedom in
  iron-based superconductors}},\ }\href
  {https://doi.org/10.1038/s41535-017-0059-y} {\bibfield  {journal} {\bibinfo
  {journal} {npj Quantum Materials}\ }\textbf {\bibinfo {volume} {2}},\
  \bibinfo {eid} {57} (\bibinfo {year} {2017})},\ \Eprint
  {https://arxiv.org/abs/1703.08622} {arXiv:1703.08622 [cond-mat.supr-con]}
  \BibitemShut {NoStop}%
\bibitem [{\citenamefont {Zhang}\ \emph {et~al.}(2021)\citenamefont {Zhang},
  \citenamefont {Devakul},\ and\ \citenamefont {Fu}}]{Zhang_2021}%
  \BibitemOpen
  \bibfield  {author} {\bibinfo {author} {\bibfnamefont {Y.}~\bibnamefont
  {Zhang}}, \bibinfo {author} {\bibfnamefont {T.}~\bibnamefont {Devakul}},\
  and\ \bibinfo {author} {\bibfnamefont {L.}~\bibnamefont {Fu}},\ }\bibfield
  {title} {\bibinfo {title} {Spin-textured chern bands in {AB}-stacked
  transition metal dichalcogenide bilayers},\ }\bibfield  {journal} {\bibinfo
  {journal} {Proceedings of the National Academy of Sciences}\ }\textbf
  {\bibinfo {volume} {118}},\ \href {https://doi.org/10.1073/pnas.2112673118}
  {10.1073/pnas.2112673118} (\bibinfo {year} {2021})\BibitemShut {NoStop}%
\bibitem [{\citenamefont {Song}\ and\ \citenamefont
  {Bernevig}(2021)}]{Song_hflTBG_2021}%
  \BibitemOpen
  \bibfield  {author} {\bibinfo {author} {\bibfnamefont {Z.-D.}\ \bibnamefont
  {Song}}\ and\ \bibinfo {author} {\bibfnamefont {B.~A.}\ \bibnamefont
  {Bernevig}},\ }\href {https://doi.org/10.48550/ARXIV.2111.05865} {\bibinfo
  {title} {Matbg as topological heavy fermion: I. exact mapping and correlated
  insulators}} (\bibinfo {year} {2021})\BibitemShut {NoStop}%
\bibitem [{\citenamefont {Ramires}\ and\ \citenamefont
  {Lado}(2021)}]{Lado_2021}%
  \BibitemOpen
  \bibfield  {author} {\bibinfo {author} {\bibfnamefont {A.}~\bibnamefont
  {Ramires}}\ and\ \bibinfo {author} {\bibfnamefont {J.~L.}\ \bibnamefont
  {Lado}},\ }\bibfield  {title} {\bibinfo {title} {Emulating heavy fermions in
  twisted trilayer graphene},\ }\href
  {https://doi.org/10.1103/PhysRevLett.127.026401} {\bibfield  {journal}
  {\bibinfo  {journal} {Phys. Rev. Lett.}\ }\textbf {\bibinfo {volume} {127}},\
  \bibinfo {pages} {026401} (\bibinfo {year} {2021})}\BibitemShut {NoStop}%
\bibitem [{\citenamefont {Dalal}\ and\ \citenamefont
  {Ruhman}(2021)}]{PhysRevResearch.3.043173}%
  \BibitemOpen
  \bibfield  {author} {\bibinfo {author} {\bibfnamefont {A.}~\bibnamefont
  {Dalal}}\ and\ \bibinfo {author} {\bibfnamefont {J.}~\bibnamefont {Ruhman}},\
  }\bibfield  {title} {\bibinfo {title} {Orbitally selective mott phase in
  electron-doped twisted transition metal-dichalcogenides: A possible
  realization of the kondo lattice model},\ }\href
  {https://doi.org/10.1103/PhysRevResearch.3.043173} {\bibfield  {journal}
  {\bibinfo  {journal} {Phys. Rev. Research}\ }\textbf {\bibinfo {volume}
  {3}},\ \bibinfo {pages} {043173} (\bibinfo {year} {2021})}\BibitemShut
  {NoStop}%
\bibitem [{\citenamefont {Kumar}\ \emph {et~al.}(2021)\citenamefont {Kumar},
  \citenamefont {Hu}, \citenamefont {MacDonald},\ and\ \citenamefont
  {Potter}}]{Potter_2021}%
  \BibitemOpen
  \bibfield  {author} {\bibinfo {author} {\bibfnamefont {A.}~\bibnamefont
  {Kumar}}, \bibinfo {author} {\bibfnamefont {N.~C.}\ \bibnamefont {Hu}},
  \bibinfo {author} {\bibfnamefont {A.~H.}\ \bibnamefont {MacDonald}},\ and\
  \bibinfo {author} {\bibfnamefont {A.~C.}\ \bibnamefont {Potter}},\ }\href
  {https://doi.org/10.48550/ARXIV.2110.11962} {\bibinfo {title} {Gate-tunable
  heavy fermion quantum criticality in a moiré kondo lattice}} (\bibinfo
  {year} {2021})\BibitemShut {NoStop}%
\bibitem [{sup()}]{supplementary}%
  \BibitemOpen
  \href@noop {} {}\bibinfo {note} {See Supplementary Material at url ... for
  details on the continuum and the tight binding Hamiltonian, the
  Schrieffer-Wolff transformation, the mean-field theory of Abrikosov
  fermions.}\BibitemShut {Stop}%
\bibitem [{\citenamefont {Schrieffer}\ and\ \citenamefont
  {Wolff}(1966)}]{PhysRev.149.491}%
  \BibitemOpen
  \bibfield  {author} {\bibinfo {author} {\bibfnamefont {J.~R.}\ \bibnamefont
  {Schrieffer}}\ and\ \bibinfo {author} {\bibfnamefont {P.~A.}\ \bibnamefont
  {Wolff}},\ }\bibfield  {title} {\bibinfo {title} {Relation between the
  anderson and kondo hamiltonians},\ }\href
  {https://doi.org/10.1103/PhysRev.149.491} {\bibfield  {journal} {\bibinfo
  {journal} {Phys. Rev.}\ }\textbf {\bibinfo {volume} {149}},\ \bibinfo {pages}
  {491} (\bibinfo {year} {1966})}\BibitemShut {NoStop}%
\bibitem [{\citenamefont {MacDonald}\ \emph {et~al.}(1988)\citenamefont
  {MacDonald}, \citenamefont {Girvin},\ and\ \citenamefont
  {Yoshioka}}]{PhysRevB.37.9753}%
  \BibitemOpen
  \bibfield  {author} {\bibinfo {author} {\bibfnamefont {A.~H.}\ \bibnamefont
  {MacDonald}}, \bibinfo {author} {\bibfnamefont {S.~M.}\ \bibnamefont
  {Girvin}},\ and\ \bibinfo {author} {\bibfnamefont {D.}~\bibnamefont
  {Yoshioka}},\ }\bibfield  {title} {\bibinfo {title} {$\frac{t}{U}$ expansion
  for the hubbard model},\ }\href {https://doi.org/10.1103/PhysRevB.37.9753}
  {\bibfield  {journal} {\bibinfo  {journal} {Phys. Rev. B}\ }\textbf {\bibinfo
  {volume} {37}},\ \bibinfo {pages} {9753} (\bibinfo {year}
  {1988})}\BibitemShut {NoStop}%
\bibitem [{\citenamefont {Coleman}\ and\ \citenamefont
  {Andrei}(1989)}]{coleman1989kondo}%
  \BibitemOpen
  \bibfield  {author} {\bibinfo {author} {\bibfnamefont {P.}~\bibnamefont
  {Coleman}}\ and\ \bibinfo {author} {\bibfnamefont {N.}~\bibnamefont
  {Andrei}},\ }\bibfield  {title} {\bibinfo {title} {Kondo-stabilised spin
  liquids and heavy fermion superconductivity},\ }\href@noop {} {\bibfield
  {journal} {\bibinfo  {journal} {Journal of Physics: Condensed Matter}\
  }\textbf {\bibinfo {volume} {1}},\ \bibinfo {pages} {4057} (\bibinfo {year}
  {1989})}\BibitemShut {NoStop}%
\bibitem [{\citenamefont {Senthil}\ \emph {et~al.}(2004)\citenamefont
  {Senthil}, \citenamefont {Vojta},\ and\ \citenamefont
  {Sachdev}}]{Senthil_2004}%
  \BibitemOpen
  \bibfield  {author} {\bibinfo {author} {\bibfnamefont {T.}~\bibnamefont
  {Senthil}}, \bibinfo {author} {\bibfnamefont {M.}~\bibnamefont {Vojta}},\
  and\ \bibinfo {author} {\bibfnamefont {S.}~\bibnamefont {Sachdev}},\
  }\bibfield  {title} {\bibinfo {title} {Weak magnetism and non-fermi liquids
  near heavy-fermion critical points},\ }\bibfield  {journal} {\bibinfo
  {journal} {Physical Review B}\ }\textbf {\bibinfo {volume} {69}},\ \href
  {https://doi.org/10.1103/physrevb.69.035111} {10.1103/physrevb.69.035111}
  (\bibinfo {year} {2004})\BibitemShut {NoStop}%
\bibitem [{\citenamefont {Pixley}\ \emph {et~al.}(2014)\citenamefont {Pixley},
  \citenamefont {Yu},\ and\ \citenamefont {Si}}]{Pixley_2014}%
  \BibitemOpen
  \bibfield  {author} {\bibinfo {author} {\bibfnamefont {J.}~\bibnamefont
  {Pixley}}, \bibinfo {author} {\bibfnamefont {R.}~\bibnamefont {Yu}},\ and\
  \bibinfo {author} {\bibfnamefont {Q.}~\bibnamefont {Si}},\ }\bibfield
  {title} {\bibinfo {title} {Quantum phases of the shastry-sutherland kondo
  lattice: Implications for the global phase diagram of heavy-fermion metals},\
  }\bibfield  {journal} {\bibinfo  {journal} {Physical Review Letters}\
  }\textbf {\bibinfo {volume} {113}},\ \href
  {https://doi.org/10.1103/physrevlett.113.176402}
  {10.1103/physrevlett.113.176402} (\bibinfo {year} {2014})\BibitemShut
  {NoStop}%
\bibitem [{\citenamefont {{Doniach}}(1977)}]{Doniach_1977}%
  \BibitemOpen
  \bibfield  {author} {\bibinfo {author} {\bibfnamefont {S.}~\bibnamefont
  {{Doniach}}},\ }\bibfield  {title} {\bibinfo {title} {{The Kondo lattice and
  weak antiferromagnetism}},\ }\href
  {https://doi.org/10.1016/0378-4363(77)90190-5} {\bibfield  {journal}
  {\bibinfo  {journal} {Physica B+C}\ }\textbf {\bibinfo {volume} {91}},\
  \bibinfo {pages} {231} (\bibinfo {year} {1977})}\BibitemShut {NoStop}%
\bibitem [{\citenamefont {Oshikawa}(2000)}]{Oshikawa_2000}%
  \BibitemOpen
  \bibfield  {author} {\bibinfo {author} {\bibfnamefont {M.}~\bibnamefont
  {Oshikawa}},\ }\bibfield  {title} {\bibinfo {title} {Topological approach to
  luttinger's theorem and the fermi surface of a kondo lattice},\ }\href
  {https://doi.org/10.1103/PhysRevLett.84.3370} {\bibfield  {journal} {\bibinfo
   {journal} {Phys. Rev. Lett.}\ }\textbf {\bibinfo {volume} {84}},\ \bibinfo
  {pages} {3370} (\bibinfo {year} {2000})}\BibitemShut {NoStop}%
\bibitem [{\citenamefont {Kane}\ and\ \citenamefont {Mele}(2005)}]{Kane_2005}%
  \BibitemOpen
  \bibfield  {author} {\bibinfo {author} {\bibfnamefont {C.~L.}\ \bibnamefont
  {Kane}}\ and\ \bibinfo {author} {\bibfnamefont {E.~J.}\ \bibnamefont
  {Mele}},\ }\bibfield  {title} {\bibinfo {title} {Quantum spin hall effect in
  graphene},\ }\bibfield  {journal} {\bibinfo  {journal} {Physical Review
  Letters}\ }\textbf {\bibinfo {volume} {95}},\ \href
  {https://doi.org/10.1103/physrevlett.95.226801}
  {10.1103/physrevlett.95.226801} (\bibinfo {year} {2005})\BibitemShut
  {NoStop}%
\bibitem [{\citenamefont {Dzero}\ \emph {et~al.}(2010)\citenamefont {Dzero},
  \citenamefont {Sun}, \citenamefont {Galitski},\ and\ \citenamefont
  {Coleman}}]{Dzero_2010}%
  \BibitemOpen
  \bibfield  {author} {\bibinfo {author} {\bibfnamefont {M.}~\bibnamefont
  {Dzero}}, \bibinfo {author} {\bibfnamefont {K.}~\bibnamefont {Sun}}, \bibinfo
  {author} {\bibfnamefont {V.}~\bibnamefont {Galitski}},\ and\ \bibinfo
  {author} {\bibfnamefont {P.}~\bibnamefont {Coleman}},\ }\bibfield  {title}
  {\bibinfo {title} {Topological kondo insulators},\ }\href
  {https://doi.org/10.1103/PhysRevLett.104.106408} {\bibfield  {journal}
  {\bibinfo  {journal} {Phys. Rev. Lett.}\ }\textbf {\bibinfo {volume} {104}},\
  \bibinfo {pages} {106408} (\bibinfo {year} {2010})}\BibitemShut {NoStop}%
\bibitem [{\citenamefont {Dzero}\ \emph {et~al.}(2016)\citenamefont {Dzero},
  \citenamefont {Xia}, \citenamefont {Galitski},\ and\ \citenamefont
  {Coleman}}]{Dzero_2016}%
  \BibitemOpen
  \bibfield  {author} {\bibinfo {author} {\bibfnamefont {M.}~\bibnamefont
  {Dzero}}, \bibinfo {author} {\bibfnamefont {J.}~\bibnamefont {Xia}}, \bibinfo
  {author} {\bibfnamefont {V.}~\bibnamefont {Galitski}},\ and\ \bibinfo
  {author} {\bibfnamefont {P.}~\bibnamefont {Coleman}},\ }\bibfield  {title}
  {\bibinfo {title} {Topological kondo insulators},\ }\href
  {https://doi.org/10.1146/annurev-conmatphys-031214-014749} {\bibfield
  {journal} {\bibinfo  {journal} {Annual Review of Condensed Matter Physics}\
  }\textbf {\bibinfo {volume} {7}},\ \bibinfo {pages} {249} (\bibinfo {year}
  {2016})}\BibitemShut {NoStop}%
\bibitem [{\citenamefont {Wolgast}\ \emph {et~al.}(2013)\citenamefont
  {Wolgast}, \citenamefont {Kurdak}, \citenamefont {Sun}, \citenamefont
  {Allen}, \citenamefont {Kim},\ and\ \citenamefont
  {Fisk}}]{PhysRevB.88.180405}%
  \BibitemOpen
  \bibfield  {author} {\bibinfo {author} {\bibfnamefont {S.}~\bibnamefont
  {Wolgast}}, \bibinfo {author} {\bibfnamefont {i.~m. c. b. u. i. e. i.~f.}\
  \bibnamefont {Kurdak}}, \bibinfo {author} {\bibfnamefont {K.}~\bibnamefont
  {Sun}}, \bibinfo {author} {\bibfnamefont {J.~W.}\ \bibnamefont {Allen}},
  \bibinfo {author} {\bibfnamefont {D.-J.}\ \bibnamefont {Kim}},\ and\ \bibinfo
  {author} {\bibfnamefont {Z.}~\bibnamefont {Fisk}},\ }\bibfield  {title}
  {\bibinfo {title} {Low-temperature surface conduction in the kondo insulator
  smb${}_{6}$},\ }\href {https://doi.org/10.1103/PhysRevB.88.180405} {\bibfield
   {journal} {\bibinfo  {journal} {Phys. Rev. B}\ }\textbf {\bibinfo {volume}
  {88}},\ \bibinfo {pages} {180405} (\bibinfo {year} {2013})}\BibitemShut
  {NoStop}%
\bibitem [{\citenamefont {Kim}\ \emph {et~al.}(2013)\citenamefont {Kim},
  \citenamefont {Thomas}, \citenamefont {Grant}, \citenamefont {Botimer},
  \citenamefont {Fisk},\ and\ \citenamefont {Xia}}]{Kim_2013}%
  \BibitemOpen
  \bibfield  {author} {\bibinfo {author} {\bibfnamefont {D.~J.}\ \bibnamefont
  {Kim}}, \bibinfo {author} {\bibfnamefont {S.}~\bibnamefont {Thomas}},
  \bibinfo {author} {\bibfnamefont {T.}~\bibnamefont {Grant}}, \bibinfo
  {author} {\bibfnamefont {J.}~\bibnamefont {Botimer}}, \bibinfo {author}
  {\bibfnamefont {Z.}~\bibnamefont {Fisk}},\ and\ \bibinfo {author}
  {\bibfnamefont {J.}~\bibnamefont {Xia}},\ }\bibfield  {title} {\bibinfo
  {title} {Surface hall effect and nonlocal transport in {SmB}6: Evidence for
  surface conduction},\ }\bibfield  {journal} {\bibinfo  {journal} {Scientific
  Reports}\ }\textbf {\bibinfo {volume} {3}},\ \href
  {https://doi.org/10.1038/srep03150} {10.1038/srep03150} (\bibinfo {year}
  {2013})\BibitemShut {NoStop}%
\bibitem [{\citenamefont {Neupane}\ \emph {et~al.}(2013)\citenamefont
  {Neupane}, \citenamefont {Alidoust}, \citenamefont {Xu}, \citenamefont
  {Kondo}, \citenamefont {Ishida}, \citenamefont {Kim}, \citenamefont {Liu},
  \citenamefont {Belopolski}, \citenamefont {Jo}, \citenamefont {Chang},
  \citenamefont {Jeng}, \citenamefont {Durakiewicz}, \citenamefont {Balicas},
  \citenamefont {Lin}, \citenamefont {Bansil}, \citenamefont {Shin},
  \citenamefont {Fisk},\ and\ \citenamefont {Hasan}}]{Neupane_2013}%
  \BibitemOpen
  \bibfield  {author} {\bibinfo {author} {\bibfnamefont {M.}~\bibnamefont
  {Neupane}}, \bibinfo {author} {\bibfnamefont {N.}~\bibnamefont {Alidoust}},
  \bibinfo {author} {\bibfnamefont {S.-Y.}\ \bibnamefont {Xu}}, \bibinfo
  {author} {\bibfnamefont {T.}~\bibnamefont {Kondo}}, \bibinfo {author}
  {\bibfnamefont {Y.}~\bibnamefont {Ishida}}, \bibinfo {author} {\bibfnamefont
  {D.~J.}\ \bibnamefont {Kim}}, \bibinfo {author} {\bibfnamefont
  {C.}~\bibnamefont {Liu}}, \bibinfo {author} {\bibfnamefont {I.}~\bibnamefont
  {Belopolski}}, \bibinfo {author} {\bibfnamefont {Y.~J.}\ \bibnamefont {Jo}},
  \bibinfo {author} {\bibfnamefont {T.-R.}\ \bibnamefont {Chang}}, \bibinfo
  {author} {\bibfnamefont {H.-T.}\ \bibnamefont {Jeng}}, \bibinfo {author}
  {\bibfnamefont {T.}~\bibnamefont {Durakiewicz}}, \bibinfo {author}
  {\bibfnamefont {L.}~\bibnamefont {Balicas}}, \bibinfo {author} {\bibfnamefont
  {H.}~\bibnamefont {Lin}}, \bibinfo {author} {\bibfnamefont {A.}~\bibnamefont
  {Bansil}}, \bibinfo {author} {\bibfnamefont {S.}~\bibnamefont {Shin}},
  \bibinfo {author} {\bibfnamefont {Z.}~\bibnamefont {Fisk}},\ and\ \bibinfo
  {author} {\bibfnamefont {M.~Z.}\ \bibnamefont {Hasan}},\ }\bibfield  {title}
  {\bibinfo {title} {Surface electronic structure of the topological
  kondo-insulator candidate correlated electron system {SmB}6},\ }\bibfield
  {journal} {\bibinfo  {journal} {Nature Communications}\ }\textbf {\bibinfo
  {volume} {4}},\ \href {https://doi.org/10.1038/ncomms3991}
  {10.1038/ncomms3991} (\bibinfo {year} {2013})\BibitemShut {NoStop}%
\bibitem [{\citenamefont {Xu}\ \emph {et~al.}(2014)\citenamefont {Xu},
  \citenamefont {Biswas}, \citenamefont {Dil}, \citenamefont {Dhaka},
  \citenamefont {Landolt}, \citenamefont {Muff}, \citenamefont {Matt},
  \citenamefont {Shi}, \citenamefont {Plumb}, \citenamefont {Radovi{\'{c}}},
  \citenamefont {Pomjakushina}, \citenamefont {Conder}, \citenamefont {Amato},
  \citenamefont {Borisenko}, \citenamefont {Yu}, \citenamefont {Weng},
  \citenamefont {Fang}, \citenamefont {Dai}, \citenamefont {Mesot},
  \citenamefont {Ding},\ and\ \citenamefont {Shi}}]{Xu_2014}%
  \BibitemOpen
  \bibfield  {author} {\bibinfo {author} {\bibfnamefont {N.}~\bibnamefont
  {Xu}}, \bibinfo {author} {\bibfnamefont {P.~K.}\ \bibnamefont {Biswas}},
  \bibinfo {author} {\bibfnamefont {J.~H.}\ \bibnamefont {Dil}}, \bibinfo
  {author} {\bibfnamefont {R.~S.}\ \bibnamefont {Dhaka}}, \bibinfo {author}
  {\bibfnamefont {G.}~\bibnamefont {Landolt}}, \bibinfo {author} {\bibfnamefont
  {S.}~\bibnamefont {Muff}}, \bibinfo {author} {\bibfnamefont {C.~E.}\
  \bibnamefont {Matt}}, \bibinfo {author} {\bibfnamefont {X.}~\bibnamefont
  {Shi}}, \bibinfo {author} {\bibfnamefont {N.~C.}\ \bibnamefont {Plumb}},
  \bibinfo {author} {\bibfnamefont {M.}~\bibnamefont {Radovi{\'{c}}}}, \bibinfo
  {author} {\bibfnamefont {E.}~\bibnamefont {Pomjakushina}}, \bibinfo {author}
  {\bibfnamefont {K.}~\bibnamefont {Conder}}, \bibinfo {author} {\bibfnamefont
  {A.}~\bibnamefont {Amato}}, \bibinfo {author} {\bibfnamefont {S.~V.}\
  \bibnamefont {Borisenko}}, \bibinfo {author} {\bibfnamefont {R.}~\bibnamefont
  {Yu}}, \bibinfo {author} {\bibfnamefont {H.-M.}\ \bibnamefont {Weng}},
  \bibinfo {author} {\bibfnamefont {Z.}~\bibnamefont {Fang}}, \bibinfo {author}
  {\bibfnamefont {X.}~\bibnamefont {Dai}}, \bibinfo {author} {\bibfnamefont
  {J.}~\bibnamefont {Mesot}}, \bibinfo {author} {\bibfnamefont
  {H.}~\bibnamefont {Ding}},\ and\ \bibinfo {author} {\bibfnamefont
  {M.}~\bibnamefont {Shi}},\ }\bibfield  {title} {\bibinfo {title} {Direct
  observation of the spin texture in {SmB}6 as evidence of the topological
  kondo insulator},\ }\bibfield  {journal} {\bibinfo  {journal} {Nature
  Communications}\ }\textbf {\bibinfo {volume} {5}},\ \href
  {https://doi.org/10.1038/ncomms5566} {10.1038/ncomms5566} (\bibinfo {year}
  {2014})\BibitemShut {NoStop}%
\bibitem [{\citenamefont {Robert}\ \emph {et~al.}(2021)\citenamefont {Robert},
  \citenamefont {Dery}, \citenamefont {Ren}, \citenamefont {Van~Tuan},
  \citenamefont {Courtade}, \citenamefont {Yang}, \citenamefont {Urbaszek},
  \citenamefont {Lagarde}, \citenamefont {Watanabe}, \citenamefont {Taniguchi},
  \citenamefont {Amand},\ and\ \citenamefont {Marie}}]{PhysRevLett.126.067403}%
  \BibitemOpen
  \bibfield  {author} {\bibinfo {author} {\bibfnamefont {C.}~\bibnamefont
  {Robert}}, \bibinfo {author} {\bibfnamefont {H.}~\bibnamefont {Dery}},
  \bibinfo {author} {\bibfnamefont {L.}~\bibnamefont {Ren}}, \bibinfo {author}
  {\bibfnamefont {D.}~\bibnamefont {Van~Tuan}}, \bibinfo {author}
  {\bibfnamefont {E.}~\bibnamefont {Courtade}}, \bibinfo {author}
  {\bibfnamefont {M.}~\bibnamefont {Yang}}, \bibinfo {author} {\bibfnamefont
  {B.}~\bibnamefont {Urbaszek}}, \bibinfo {author} {\bibfnamefont
  {D.}~\bibnamefont {Lagarde}}, \bibinfo {author} {\bibfnamefont
  {K.}~\bibnamefont {Watanabe}}, \bibinfo {author} {\bibfnamefont
  {T.}~\bibnamefont {Taniguchi}}, \bibinfo {author} {\bibfnamefont
  {T.}~\bibnamefont {Amand}},\ and\ \bibinfo {author} {\bibfnamefont
  {X.}~\bibnamefont {Marie}},\ }\bibfield  {title} {\bibinfo {title}
  {Measurement of conduction and valence bands $g$-factors in a transition
  metal dichalcogenide monolayer},\ }\href
  {https://doi.org/10.1103/PhysRevLett.126.067403} {\bibfield  {journal}
  {\bibinfo  {journal} {Phys. Rev. Lett.}\ }\textbf {\bibinfo {volume} {126}},\
  \bibinfo {pages} {067403} (\bibinfo {year} {2021})}\BibitemShut {NoStop}%
\bibitem [{\citenamefont {Ong}(1991)}]{PhysRevB.43.193}%
  \BibitemOpen
  \bibfield  {author} {\bibinfo {author} {\bibfnamefont {N.~P.}\ \bibnamefont
  {Ong}},\ }\bibfield  {title} {\bibinfo {title} {Geometric interpretation of
  the weak-field hall conductivity in two-dimensional metals with arbitrary
  fermi surface},\ }\href {https://doi.org/10.1103/PhysRevB.43.193} {\bibfield
  {journal} {\bibinfo  {journal} {Phys. Rev. B}\ }\textbf {\bibinfo {volume}
  {43}},\ \bibinfo {pages} {193} (\bibinfo {year} {1991})}\BibitemShut
  {NoStop}%
\bibitem [{\citenamefont {Haldane}(2004)}]{Haldane_2004}%
  \BibitemOpen
  \bibfield  {author} {\bibinfo {author} {\bibfnamefont {F.~D.~M.}\
  \bibnamefont {Haldane}},\ }\bibfield  {title} {\bibinfo {title} {Berry
  curvature on the fermi surface: Anomalous hall effect as a topological
  fermi-liquid property},\ }\bibfield  {journal} {\bibinfo  {journal} {Physical
  Review Letters}\ }\textbf {\bibinfo {volume} {93}},\ \href
  {https://doi.org/10.1103/physrevlett.93.206602}
  {10.1103/physrevlett.93.206602} (\bibinfo {year} {2004})\BibitemShut
  {NoStop}%
\bibitem [{\citenamefont {Chowdhury}\ \emph {et~al.}(2018)\citenamefont
  {Chowdhury}, \citenamefont {Sodemann},\ and\ \citenamefont
  {Senthil}}]{Chowdhury_2018}%
  \BibitemOpen
  \bibfield  {author} {\bibinfo {author} {\bibfnamefont {D.}~\bibnamefont
  {Chowdhury}}, \bibinfo {author} {\bibfnamefont {I.}~\bibnamefont
  {Sodemann}},\ and\ \bibinfo {author} {\bibfnamefont {T.}~\bibnamefont
  {Senthil}},\ }\bibfield  {title} {\bibinfo {title} {Mixed-valence insulators
  with neutral fermi surfaces},\ }\bibfield  {journal} {\bibinfo  {journal}
  {Nature Communications}\ }\textbf {\bibinfo {volume} {9}},\ \href
  {https://doi.org/10.1038/s41467-018-04163-2} {10.1038/s41467-018-04163-2}
  (\bibinfo {year} {2018})\BibitemShut {NoStop}%
\bibitem [{\citenamefont {Ioffe}\ and\ \citenamefont
  {Larkin}(1989)}]{Larkin_Ioffe_1989}%
  \BibitemOpen
  \bibfield  {author} {\bibinfo {author} {\bibfnamefont {L.~B.}\ \bibnamefont
  {Ioffe}}\ and\ \bibinfo {author} {\bibfnamefont {A.~I.}\ \bibnamefont
  {Larkin}},\ }\bibfield  {title} {\bibinfo {title} {Gapless fermions and gauge
  fields in dielectrics},\ }\href {https://doi.org/10.1103/PhysRevB.39.8988}
  {\bibfield  {journal} {\bibinfo  {journal} {Phys. Rev. B}\ }\textbf {\bibinfo
  {volume} {39}},\ \bibinfo {pages} {8988} (\bibinfo {year}
  {1989})}\BibitemShut {NoStop}%
\bibitem [{\citenamefont {Paschen}\ \emph {et~al.}(2004)\citenamefont
  {Paschen}, \citenamefont {Lühmann}, \citenamefont {Wirth}, \citenamefont
  {Gegenwart}, \citenamefont {Trovarelli}, \citenamefont {Geibel},
  \citenamefont {Steglich}, \citenamefont {Coleman},\ and\ \citenamefont
  {Si}}]{Paschen_2004}%
  \BibitemOpen
  \bibfield  {author} {\bibinfo {author} {\bibfnamefont {S.}~\bibnamefont
  {Paschen}}, \bibinfo {author} {\bibfnamefont {T.}~\bibnamefont {Lühmann}},
  \bibinfo {author} {\bibfnamefont {S.}~\bibnamefont {Wirth}}, \bibinfo
  {author} {\bibfnamefont {P.}~\bibnamefont {Gegenwart}}, \bibinfo {author}
  {\bibfnamefont {O.}~\bibnamefont {Trovarelli}}, \bibinfo {author}
  {\bibfnamefont {C.}~\bibnamefont {Geibel}}, \bibinfo {author} {\bibfnamefont
  {F.}~\bibnamefont {Steglich}}, \bibinfo {author} {\bibfnamefont
  {P.}~\bibnamefont {Coleman}},\ and\ \bibinfo {author} {\bibfnamefont
  {Q.}~\bibnamefont {Si}},\ }\bibfield  {title} {\bibinfo {title} {Hall-effect
  evolution across a heavy-fermion quantum critical point},\ }\href
  {https://doi.org/10.1038/nature03129} {\bibfield  {journal} {\bibinfo
  {journal} {Nature}\ }\textbf {\bibinfo {volume} {432}},\ \bibinfo {pages}
  {881} (\bibinfo {year} {2004})}\BibitemShut {NoStop}%
\bibitem [{\citenamefont {Friedemann}\ \emph
  {et~al.}(2010{\natexlab{a}})\citenamefont {Friedemann}, \citenamefont
  {Oeschler}, \citenamefont {Wirth}, \citenamefont {Krellner}, \citenamefont
  {Geibel}, \citenamefont {Steglich}, \citenamefont {Paschen}, \citenamefont
  {Kirchner},\ and\ \citenamefont {Si}}]{Friedemann_pnas_2010}%
  \BibitemOpen
  \bibfield  {author} {\bibinfo {author} {\bibfnamefont {S.}~\bibnamefont
  {Friedemann}}, \bibinfo {author} {\bibfnamefont {N.}~\bibnamefont
  {Oeschler}}, \bibinfo {author} {\bibfnamefont {S.}~\bibnamefont {Wirth}},
  \bibinfo {author} {\bibfnamefont {C.}~\bibnamefont {Krellner}}, \bibinfo
  {author} {\bibfnamefont {C.}~\bibnamefont {Geibel}}, \bibinfo {author}
  {\bibfnamefont {F.}~\bibnamefont {Steglich}}, \bibinfo {author}
  {\bibfnamefont {S.}~\bibnamefont {Paschen}}, \bibinfo {author} {\bibfnamefont
  {S.}~\bibnamefont {Kirchner}},\ and\ \bibinfo {author} {\bibfnamefont
  {Q.}~\bibnamefont {Si}},\ }\bibfield  {title} {\bibinfo {title}
  {Fermi-surface collapse and dynamical scaling near a quantum-critical
  point},\ }\href {https://doi.org/10.1073/pnas.1009202107} {\bibfield
  {journal} {\bibinfo  {journal} {Proceedings of the National Academy of
  Sciences}\ }\textbf {\bibinfo {volume} {107}},\ \bibinfo {pages} {14547}
  (\bibinfo {year} {2010}{\natexlab{a}})},\ \Eprint
  {https://arxiv.org/abs/https://www.pnas.org/doi/pdf/10.1073/pnas.1009202107}
  {https://www.pnas.org/doi/pdf/10.1073/pnas.1009202107} \BibitemShut {NoStop}%
\bibitem [{\citenamefont {Friedemann}\ \emph
  {et~al.}(2010{\natexlab{b}})\citenamefont {Friedemann}, \citenamefont
  {Wirth}, \citenamefont {Oeschler}, \citenamefont {Krellner}, \citenamefont
  {Geibel}, \citenamefont {Steglich}, \citenamefont {MaQuilon}, \citenamefont
  {Fisk}, \citenamefont {Paschen},\ and\ \citenamefont
  {Zwicknagl}}]{PhysRevB.82.035103}%
  \BibitemOpen
  \bibfield  {author} {\bibinfo {author} {\bibfnamefont {S.}~\bibnamefont
  {Friedemann}}, \bibinfo {author} {\bibfnamefont {S.}~\bibnamefont {Wirth}},
  \bibinfo {author} {\bibfnamefont {N.}~\bibnamefont {Oeschler}}, \bibinfo
  {author} {\bibfnamefont {C.}~\bibnamefont {Krellner}}, \bibinfo {author}
  {\bibfnamefont {C.}~\bibnamefont {Geibel}}, \bibinfo {author} {\bibfnamefont
  {F.}~\bibnamefont {Steglich}}, \bibinfo {author} {\bibfnamefont
  {S.}~\bibnamefont {MaQuilon}}, \bibinfo {author} {\bibfnamefont
  {Z.}~\bibnamefont {Fisk}}, \bibinfo {author} {\bibfnamefont {S.}~\bibnamefont
  {Paschen}},\ and\ \bibinfo {author} {\bibfnamefont {G.}~\bibnamefont
  {Zwicknagl}},\ }\bibfield  {title} {\bibinfo {title} {Hall effect
  measurements and electronic structure calculations on
  ${\text{ybrh}}_{2}{\text{si}}_{2}$ and its reference compounds
  ${\text{lurh}}_{2}{\text{si}}_{2}$ and ${\text{ybir}}_{2}{\text{si}}_{2}$},\
  }\href {https://doi.org/10.1103/PhysRevB.82.035103} {\bibfield  {journal}
  {\bibinfo  {journal} {Phys. Rev. B}\ }\textbf {\bibinfo {volume} {82}},\
  \bibinfo {pages} {035103} (\bibinfo {year} {2010}{\natexlab{b}})}\BibitemShut
  {NoStop}%
\bibitem [{\citenamefont {Ding}\ \emph {et~al.}(2015)\citenamefont {Ding},
  \citenamefont {Grefe}, \citenamefont {Paschen},\ and\ \citenamefont
  {Si}}]{Ding_2015}%
  \BibitemOpen
  \bibfield  {author} {\bibinfo {author} {\bibfnamefont {W.}~\bibnamefont
  {Ding}}, \bibinfo {author} {\bibfnamefont {S.}~\bibnamefont {Grefe}},
  \bibinfo {author} {\bibfnamefont {S.}~\bibnamefont {Paschen}},\ and\ \bibinfo
  {author} {\bibfnamefont {Q.}~\bibnamefont {Si}},\ }\href
  {https://doi.org/10.48550/ARXIV.1507.07328} {\bibinfo {title} {Anomalous hall
  effect and quantum criticality in geometrically frustrated heavy fermion
  metals}} (\bibinfo {year} {2015})\BibitemShut {NoStop}%
\bibitem [{\citenamefont {Zhao}\ \emph
  {et~al.}(2022{\natexlab{b}})\citenamefont {Zhao}, \citenamefont {Shen},
  \citenamefont {Tao}, \citenamefont {Han}, \citenamefont {Kang}, \citenamefont
  {Watanabe}, \citenamefont {Taniguchi}, \citenamefont {Mak},\ and\
  \citenamefont {Shan}}]{Zhao_2022}%
  \BibitemOpen
  \bibfield  {author} {\bibinfo {author} {\bibfnamefont {W.}~\bibnamefont
  {Zhao}}, \bibinfo {author} {\bibfnamefont {B.}~\bibnamefont {Shen}}, \bibinfo
  {author} {\bibfnamefont {Z.}~\bibnamefont {Tao}}, \bibinfo {author}
  {\bibfnamefont {Z.}~\bibnamefont {Han}}, \bibinfo {author} {\bibfnamefont
  {K.}~\bibnamefont {Kang}}, \bibinfo {author} {\bibfnamefont {K.}~\bibnamefont
  {Watanabe}}, \bibinfo {author} {\bibfnamefont {T.}~\bibnamefont {Taniguchi}},
  \bibinfo {author} {\bibfnamefont {K.~F.}\ \bibnamefont {Mak}},\ and\ \bibinfo
  {author} {\bibfnamefont {J.}~\bibnamefont {Shan}},\ }\href
  {https://doi.org/10.48550/ARXIV.2211.00263} {\bibinfo {title} {Gate-tunable
  heavy fermions in a moiré kondo lattice}} (\bibinfo {year}
  {2022}{\natexlab{b}})\BibitemShut {NoStop}%
\bibitem [{\citenamefont {Cr{\'{e} }pel}\ and\ \citenamefont
  {Fu}(2022)}]{Cr_pel_2022}%
  \BibitemOpen
  \bibfield  {author} {\bibinfo {author} {\bibfnamefont {V.}~\bibnamefont
  {Cr{\'{e} }pel}}\ and\ \bibinfo {author} {\bibfnamefont {L.}~\bibnamefont
  {Fu}},\ }\bibfield  {title} {\bibinfo {title} {Spin-triplet superconductivity
  from excitonic effect in doped insulators},\ }\bibfield  {journal} {\bibinfo
  {journal} {Proceedings of the National Academy of Sciences}\ }\textbf
  {\bibinfo {volume} {119}},\ \href {https://doi.org/10.1073/pnas.2117735119}
  {10.1073/pnas.2117735119} (\bibinfo {year} {2022})\BibitemShut {NoStop}%
\end{thebibliography}%


\onecolumngrid
\newpage
\makeatletter

\begin{center}
\textbf{\large Supplementary material for `` \@title ''} \\[10pt]
Daniele Guerci$^\mathsection$, Jie Wang$^\mathsection$, Jiawei Zhang, Jennifer Cano, J. H. Pixley and Andrew Millis \\ 
$^\mathsection$These authors contributed equally. \\
\end{center}
\vspace{20pt}

\setcounter{figure}{0}
\setcounter{section}{0}
\setcounter{equation}{0}

\renewcommand{\thefigure}{S\@arabic\c@figure}
\makeatother

\onecolumngrid
\appendix


These supplementary materials contain the details of analytic calculations as well as additional numerical details supporting the results presented in the main text.

\section{The Kondo lattice Hamiltonian}

In this section we detail the connection between the tight-binding model and the Schrieffer-Wolff transformation to obtain the Kondo lattice Hamiltonian given in the main text.

\subsection{The tight-binding Hamiltonian}
Assuming $W$-layer weakly correlated the tight-binding Hamiltonian that describes the low-energy properties of the system reads $H=H_0+H_W+H_t$: 
\begin{equation}
\label{H_tb}
\begin{split}
    H_0=&-\frac{\Delta}{2} (N_{Mo}-N_W)+U_{}\sum_{\br\in Mo}n_{\br\uparrow}n_{\br\downarrow}\\
    H_W=&- t_{W}\sum_{\langle \br,\br'\rangle\in W} e^{-i\nu_{\br,\br'}2\pi \sigma/3} c^\dagger_{\br\sigma} c_{\br'\sigma},\\
    H_t=&-t_{Mo} \sum_{\langle \br,\br'\rangle\in Mo} f^\dagger_{\br\sigma} f_{\br'\sigma}-t_h\sum_{\langle\br,\br'\rangle}f^\dagger_{\br}c_{\br'}.
\end{split}
\end{equation}
We have introduced the operator $f_{\br\sigma}$ for the $Mo$ and $c_{\br\sigma}$ for $W$-layer, respectively.   
We notice that differently from $H_0$ and $H_W$ the contribution $H_t$ changes the local configuration in the $Mo$ layer.    

\subsection{The Schrieffer-Wolff transformation}
\begin{figure}
    \centering
    \includegraphics[width=0.5\textwidth]{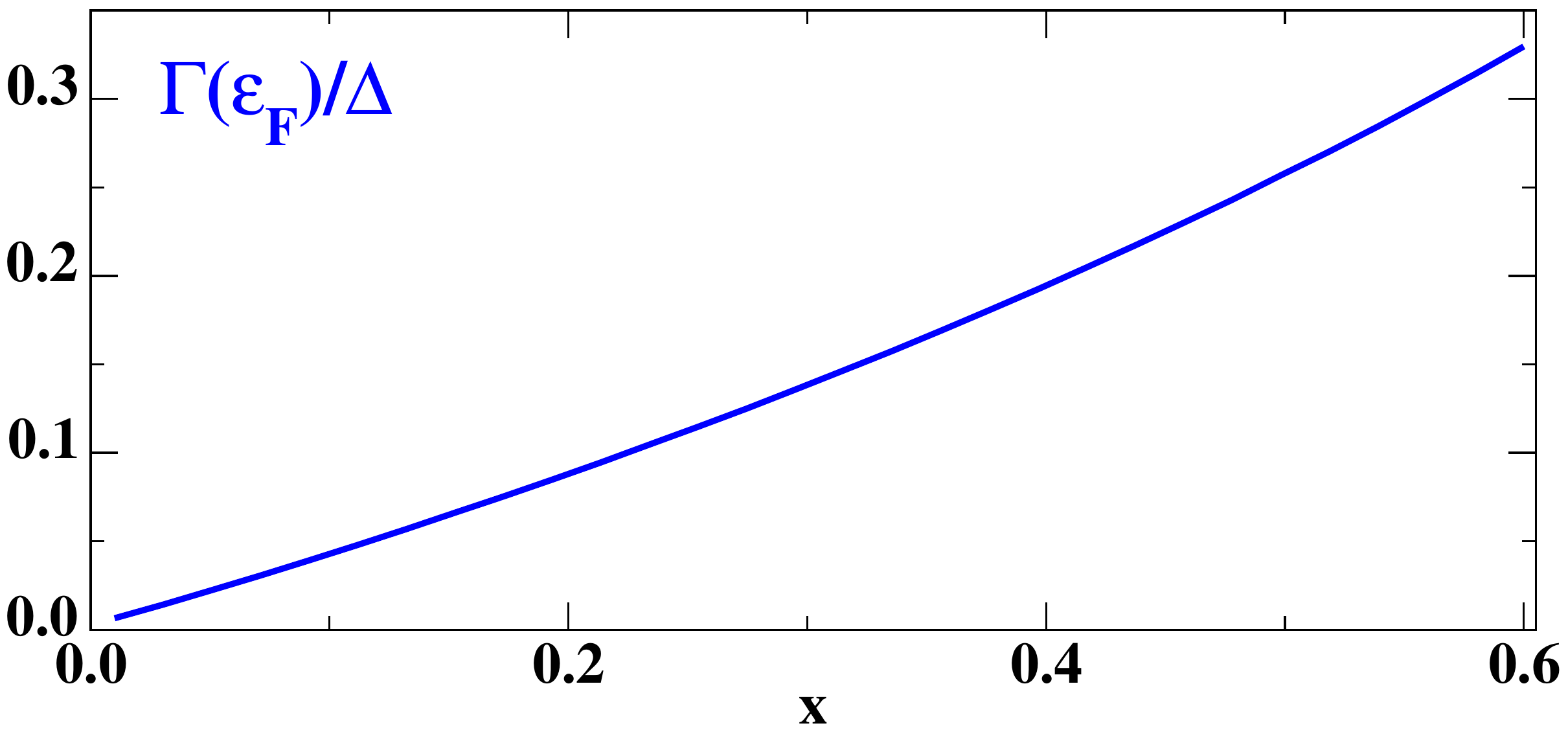}
    \caption{\textbf{Hybridization function}. Evolution of $\Gamma(\epsilon_F)/\Delta$ as a function of the filling $x$ of the conduction band.}
    \label{fig:hybrd}
\end{figure}

We assume that $U$ is the largest energy scale, $U\gg t_{Mo}$, and $\Gamma(\epsilon_F)/\Delta<1$ where $\Gamma(\epsilon_F)$ is the hybridization function:
 \begin{equation}
     \Gamma(\epsilon_F)=\pi t^2_h\rho(\epsilon_F)\langle V^*_\bk V_\bk\rangle_{FS}=\pi t^2_h\left(\frac{3\sqrt{3}a^2_M}{8\pi^2}\right)\oint_{FS} dk_t\frac{V^*_\bk V_\bk}{|\nabla_\bk \epsilon_{\bk\sigma}|}.
 \end{equation}
Fig.~\ref{fig:hybrd} show the evolution of $\Gamma(\epsilon_F)$ as a function of the filling $x$ for the value of $t^2_h/\Delta$ corresponding to $J_K/t_W=1$ value used for the numerical calculations shown in the manuscript. Within these assumptions valence fluctuations in $Mo$ layer are suppressed. We now observe that the tunneling $H_t$ can be decomposed as:
\begin{equation}
    H_t=\sum_{ q=-1}^1\sum_{ d=-1}^1 T_{ q, d},
\end{equation}
where $T_{ q, d}$ gathers all tunneling events that change the number of electrons unbalance between $Mo$ and $W$ by $ q$ and the double occupancies in $Mo$ layer by $d$. We list the $T_{q,d}$ operators below: \begin{equation}
    \begin{split}
        &T_{+1,+1}=-t_h\sum_{\br\in Mo}\sum_\sigma\sum_{j=1}^3 n_{\br\bar\sigma}f^\dagger_{\br\sigma}c_{\br+\bm\delta_j\sigma},\\
        &T_{+1,0}=-t_h\sum_{\br\in Mo}\sum_\sigma\sum_{j=1}^3 h_{\br\bar\sigma}f^\dagger_{\br\sigma}c_{\br+\bm\delta_j\sigma},\\
        &T_{0,+1}=-t_{Mo}\sum_{\br\in Mo}\sum_\sigma\sum_{j=1}^3\left[ n_{\br\bar\sigma}f^\dagger_{\br\sigma}f_{\br+\bm\gamma_j\sigma}h_{\br+\bm\gamma_j\bar\sigma}+ n_{\br+\bm\gamma_j\bar\sigma}f^\dagger_{\br+\bm\gamma_j\sigma}f_{\br\sigma}h_{\br\bar\sigma} \right],\\
        &T_{0,0}=-t_{Mo}\sum_{\br\in Mo}\sum_\sigma\sum_{j=1}^3\left[ n_{\br\bar\sigma}f^\dagger_{\br\sigma}f_{\br+\bm\gamma_j\sigma}n_{\br+\bm\gamma_j\bar\sigma}+ h_{\br\bar\sigma}f^\dagger_{\br\sigma}f_{\br+\bm\gamma_j\sigma}h_{\br+\bm\gamma_j\bar\sigma} + h.c. \right],
    \end{split}
\end{equation}
we notice that $T_{-q,-d}=T^\dagger_{q,d}$, $n_{\br\sigma}=f^\dagger_{\br\sigma} f_{\br\sigma}$ and $h_{\br\sigma}=1-n_{\br\sigma}$. Moreover, the terms $T_{-1,+1}$ and $T_{+1,-1}$ vanish.  
Before moving on, we notice that the intralayer hopping $H_{Mo}$ does not change the charge imbalance $N_{Mo}-N_{W}$ ($q=0$), i.e. $H_{Mo}$ commutes with $N_{Mo}-N_W$ ($[N_{Mo}-N_{W},H_{Mo}]=0$). Moreover, $[N_{Mo}-N_W,T_{ q, d}]=2q T_{ q, d}$ and
\begin{equation}
    \sum_{\br\in Mo}[n_{\br\uparrow}n_{\br\downarrow},T_{ q, d}]= d T_{q, d}.
\end{equation}
We seek  an unitary transformation $U=\exp(-iS)$ which eliminates hops between states with different numbers of doubly occupied sites and interlayer charge imbalance~\cite{PhysRev.149.491,PhysRevB.37.9753,Cr_pel_2022}:
\begin{equation}
    \bar H\equiv e^{iS} H e^{-iS}=H+[iS,H]+\frac{[iS,[iS,H]]}{2!}+\cdots,
\end{equation}
where we applied the Baker-Campbell-Haussdorf formula. We now notice that to the lowest order in the expansion we have: 
\begin{equation}
\label{SW_1st}
    [H_0,iS]=\sum_{q,d}^{(q,d)\neq0}T_{q,d},
\end{equation}
where in previous sum we exclude the term with $q=0$ and $d=0$ in shorthand notation $(q,d)\neq0$. We readily realize that the solution of Eq.~\eqn{SW_1st} reads: 
\begin{equation}
    S=-i\sum_{q,d}^{(q,d)\neq0}\frac{T_{q,d}}{d U-q\Delta}.
\end{equation}  
From the latter expression we find:
\begin{equation}
\begin{split}    
    \bar H=& H_0+T_{0,0}+H_W+\frac{1}{2}\sum_{q,d}^{(q,d)\neq0}\sum_{q',d'}^{(q',d')\neq0}\frac{[T_{q,d},T_{q',d'}]}{-q\Delta + dU}+\sum_{q,d}^{(q,d)\neq0}\frac{[T_{q,d},H_W+T_{0,0}]}{-q\Delta + dU}+O([H_W+T_{0,0}]S^2).
\end{split}
\end{equation}
Projecting the model in the low-energy subspace with one-electron in $Mo$ layer, i.e. $n_{\br}=1$ for $\br\in Mo$, we find that the latter term vanishes. 
Furthermore, the projection constraints $q'=-q$ and $d'=-d$:  
\begin{equation}
\label{sm:Hsd_0}
    \bar H=- t_{W}\sum_{\langle \br,\br'\rangle\in W} e^{-i\nu_{\br,\br'}2\pi s_\sigma/3} c^\dagger_{\br\sigma} c_{\br'\sigma}-\mu N_W+\frac{1}{2}\sum_{q,d}^{(q,d)\neq0}\frac{[T_{q,d},T_{-q,-d}]}{ dU-q\Delta},
\end{equation}
where the constant energy term $\Delta N_{W}/2$ has been absorbed in the chemical potential shift and $T_{0,0}$ vanishes in the low-energy subspace. 
By performing straightforward calculations we find that the commutator in Eq.~\eqn{sm:Hsd_0} gives: 
\begin{equation}
\begin{split}
    \frac{1}{2}\sum_{q,d}^{(q,d)\neq0}\frac{[T_{q,d},T_{-q,-d}]}{ dU-q\Delta}=&t^2_h\left(\frac{1}{\Delta}+\frac{1}{U-\Delta}\right)\sum_{\br\in Mo}\sum_{jl=1}^3\bm S_\br\cdot c^\dagger_{\br+\bm\delta_j \sigma} \,\bm \sigma_{\sigma\sigma'}\, c_{\br+\bm\delta_l\sigma'}+\frac{4t^2_{Mo}}{U}\sum_{\langle\br,\br'\rangle\in A}\bm S_\br \cdot\bm S_{\br'}\\
    &+t^2_h\left(\frac{1}{\Delta}-\frac{1}{U-\Delta}\right)\sum_\bk |V_\bk|^2 c^\dagger_\bk c_\bk. 
\end{split}
\end{equation}
Expanding close to the bottom of the conduction band [$\kappa(\kappa')$ for $\uparrow(\downarrow)$] $|V_{\kappa/\kappa'+\bk}|^2\simeq 9(a_Mk)^2/4$ we realize that the latter terms can be simply accounted by a redefinition of the bare mass $m_W\to m_W/[1+(t_h/\Delta-t_h/(U-\Delta))t_h/(2t_W)]$.
Finally, we find the effective spin-fermion Hamiltonian:
\begin{equation}
\label{sm:Hsd}
    \bar H = \sum_{\bk}  \xi_{\bk\sigma}c^\dagger_{\bk\sigma} c_{\bk\sigma}+ J_H\sum_{\langle\br,\br'\rangle\in A}\bm S_\br \cdot\bm S_{\br'}+\frac{1}{2N}\sum_{\br\in \text{Mo}}\sum_{\bk,\bp} e^{-i(\bk-\bp)\cdot\br}
    J_{\bk,\bp}
    \bm S_\br\cdot c^\dagger_{\bk \sigma} \,\bm \sigma_{\sigma\sigma'}\, c_{\bp\sigma'},
\end{equation}
where $J_H=4t^2_{Mo}/U$, $J_{\bk,\bp}=J_K V^*_\bk V_\bp$, $J_K=2t^2_h[1/\Delta+1/(U-\Delta)]$, $\xi_{\bk\sigma}=\epsilon_{\bk\sigma}-\epsilon_F$, $\epsilon_{\bk\sigma}$ is the electron dispersion and $\epsilon_F$ fixes the number of electron in the conduction band. The dispersion relation reads $\epsilon_{\bk\sigma}=-2t_W\sum^3_{j=1}\cos(\bk\cdot\bm a_i+2\pi s_\sigma/3)$, $\bm a_1=\sqrt{3}a_{M}(1,0)$, $\bm a_{2,3}=\sqrt{3}a_{M}\left(-1/2,\pm\sqrt{3}/2\right)$ are the lattice vectors with $a_{M}=5$nm the moir\'e cell lattice constant. The form factor is $V_\bk=\sum_{j=1}^3 e^{i\bk\cdot\bm\delta_j}$. Despite $U$ is large and $J_H$ small the particular form of the exchange interaction $J_{\bk,\bp}=J_K V^*_\bk V_\bp$ gives rise to a non-trivial competition between a low-density magnetic phase and a paramagnetic heavy Fermi liquid. We observe that in the following we will measure the value of $\Delta$ with respect to the bottom of conduction band $-6t_W$. 

\subsubsection{On the higher-order corrections}

Higher order corrections in the expansion lower the symmetry of the spin interaction in the $Mo$-layer. This can be simply realized noticing that the phase factor in the Hamiltonian $H_W$ lower the spin-symmetry of the model to $U(1)$ around $z$. Including higher-order hopping processes mediated by $t_h$ gives an XXZ spin model with Dzyaloshinskii–Moriya interactions:   
\begin{equation}
    H_S=J_H\sum_{\langle\br,\br'\rangle\in \text{Mo}}\left(S^z_\br S^z_{\br'}+\gamma S^+_\br S^-_{\br'}+h.c.\right)+D\sum_{\langle\br,\br'\rangle\in \text{Mo}}\left(\bm S_\br\times \bm S_{\br'}\right)_z,
\end{equation}
we refer to Ref.~\cite{Devakul_magic_2021} for additional details. For simplicity we consider the isotropic limit $\gamma=1/2$ and $D=0$ in our calculations.

\section{The mean-field approach}

In this section we detail the mean-field of Abrikosov fermions discussed in the main text. We perform the decomposition of the spin-$1/2$ into spinons $\bm S_\br=\chi^\dagger_{\br\alpha}\bm\sigma_{\alpha\beta}\chi_{\br\beta}/2$ and performing the mean-field decomposition in the magnetic and excitonic channels we obtain the mean-field Hamiltonian: 
\begin{equation}
\label{mean_field_Ham}
\begin{split}    
    \bar H_{\rm mf}=&\lambda\sum_{\bk} \chi^\dagger_\bk \chi_\bk+\frac{h^z}{2}\sum_{\bk} \chi^\dagger_\bk \sigma^z \chi_\bk +\frac{h^\parallel}{2}\sum_{\bk}\left[ \chi^\dagger_{\bk+\bQ} \sigma^- \chi_{\bk}+h.c. \right]+\sum_{\bk\sigma}\xi_{\bk\sigma}c^\dagger_{\bk\sigma}c_{\bk\sigma}\\
    &+\frac{ M^z}{2}\sum_\bk J_{\bk,\bk} c^\dagger_\bk\sigma^z c_\bk+\frac{M^\parallel}{2}\sum_\bk\left[ J_{\bk+\bQ,\bk} c^\dagger_{\bk+\bQ}\sigma^- c_\bk+h.c.\right]\\
    &+J_K\sum_{\bk}\left[V^*_\bk\Phi^* c^\dagger_{\bk} \chi_{\bk}+h.c.\right].
\end{split}
\end{equation} 
In the previous expression $M^z$ and $M^\parallel$ are the out-of-plane and in-plane components of the magnetization of the local moments, the magnetic field $h^z$ and $h^\parallel$ are given by: 
\begin{equation}
\begin{split}
    &h^z=6M^z+\sum_{\bp}J_{\bp,\bp}{\langle c^\dagger_{\bp}\sigma^z c_{\bp}\rangle}/N,\\
    & h^\parallel=-3M^\parallel+\sum_{\bp}\left(J_{\bp,\bp+\bQ}{\langle c^\dagger_{\bp}\sigma^+ c_{\bp+\bQ}\rangle}+h.c.\right)/(2N).
\end{split}
\end{equation}
The excitonic order parameter reads: 
\begin{equation}
    \Phi=-\sum_{\bk}V^*_{\bk}\langle c^\dagger_{\bk} \chi_{\bk}\rangle/(2N).
\end{equation}
The mean-field free-energy reads: 
\begin{equation}
\begin{split}
\label{meafield_free}
    F_{\rm mf}=&-k_BT\sum_{\bk}\sum_\lambda\log\left[1+e^{-E_{\bk\lambda}/k_BT}\right]+2J_K N |\Phi |^2+\frac{3J_H N}{2} {M^\parallel}^2\\
    &-3J_H N {M^z}^2-J_K M^z N m^z-J_K N M^\parallel m^\parallel-\lambda N+N \mu x,
\end{split}
\end{equation}
where $E_{\bk\lambda}$ is the mean-field dispersion relation. The  mean-field solution is obtained by minimizing $F_{\rm mf}$ with respect to the variational parameters $M^z, M^\parallel, \Phi, m^z, m^\parallel$. The Lagrange multiplier $\lambda$ imposes the Gutzwiller constraint $\sum_\bk\langle \chi^\dagger_\bk\chi_\bk\rangle/N=1$, while the chemical potential $\mu$ fixes the number of particle in conduction band $\sum_\bk \langle c^\dagger_{\bk}c_{\bk}\rangle/N=x$. Taking the saddle point of Eq.~\eqn{meafield_free} with respect to the variational parameters gives a set of self-consistency equations that are solved by find-root algorithm. 

\subsection{Energetics of the HFL and AFM states}
\label{energetics}

In this section we detail the weak coupling expansion to determines the characteristic energy scales of the AFM and HFL phases that are the RKKY energy and the Kondo temperature, respectively. 
We will also introduce the effective model describing the quasiparticle excitations in the two different regimes.   

\subsubsection{AFM}
In the magnetic regime the local moments $\bm S_\br$ form a 120$^\circ$ AFM order with $\langle\bm S_\br\rangle=\left(M^\parallel\cos\bQ\cdot\br,M^\parallel \sin\bQ\cdot\br,M^z\right)$, $\bQ=\kappa-\kappa'$,  correspondingly the conduction electron Hamiltonian reads: 
\begin{equation}
\begin{split}
\label{AFM_lattice}
    \bar H^c_{\rm mf} =&\sum_\bk \xi_{\bk \sigma}c^\dagger_{\bk\sigma}c_{\bk\sigma}+\frac{J_K M^z}{2}\sum_\bk |V_{\bk }|^2 c^\dagger_\bk\sigma^z c_\bk+\frac{J_K M^\parallel}{2}\sum_{\bk}\left[V^*_{\bk }V_{\bk+\bQ }c^\dagger_\bk\sigma^+ c_{\bk+\bQ}+h.c.\right].
\end{split}
\end{equation}
In the limit of low-doping the Fermi surface is a small electron pocket around $\kappa(\uparrow)$ and $\kappa'(\downarrow)$ that are folded into the origin of the magnetic Brillouin zone $\gamma_m$, depicted in Fig.~\ref{fig:fig_continuum}(b). Expanding close to quadratic order around $\gamma_m$ and keeping only the two lowest energy bands we find the continuum model: 
\begin{equation}
\label{conduction_electrons}
    \bar{\mathcal H}^c_{\rm mf}=\sum_{\bk}c^\dagger_{\bk}\left[\left(\frac{\hbar^2 k^2}{2m_W}-\epsilon_F\right)\sigma_0+\frac{9J_Ka^2_M }{8}\sum_{a}d_a(\bk)\sigma_a \right]c_\bk,
\end{equation}
where $d_z(\bk)=M^z k^2$, $d_x(\bk)=-M^\parallel(k^2_x-k^2_y)$ and $d_y(\bk)=2M^\parallel k_xk_y$.
We readily realize that due to the SOC term the spin is no longer a good quantum number, the eigenstates $\ket{u_{\bk\lambda}}$ are labeled by $\lambda=\pm$ and the corresponding eigenvalues are $\epsilon_{\bk\lambda}=(\hbar^2k^2/2m_W-\epsilon_F)+\lambda 9J_Ka^2_Mk^2|M|/8$ with $|M|=\sqrt{(M^z)^2+(M^\parallel)^2}$.
We observe that the theory is O$(3)$ invariant under rotation of the magnetization $M^a\to R_{ab}M^b$. The resulting stabilization energy does not depend on the orientation of the local moments. 
The Fermi momentum obtained by setting $\epsilon_{\bk\lambda}=0$ reads $k^{\pm}_F=\sqrt{2m_\lambda\epsilon_F/\hbar^2}$ where $m_\lambda=m_W/[1+\lambda (J_K/8t_W)]$ and the Fermi energy is $\epsilon_F\simeq x/\left(\sum_\lambda \rho_\lambda\right)$ with $\rho_\lambda=\rho_0 m_\lambda/m_W$. We notice that in the Kondo regime $J_K$ is smaller than the bandwidth of conduction electrons $9t_W$ so that the mass is always positive. The kinetic energy variation with respect to the normal state reads:
\begin{equation}
    \delta \epsilon^c_{\rm mf}=\epsilon^c_{\rm mf}-\epsilon^c_{J_K=0}=-\rho_0\bar J^2_K|M|^2/2,
\end{equation}
where $\bar J_K$ is the average over the FS of the Kondo exchange.
We conclude that the total energy per site in the magnetic regime is given by: 
\begin{equation}
    \epsilon_{\rm mf}=-3J_H |M|^2/2-\rho_0\bar J^2_K|M|^2/2.
\end{equation}
The energy gain from the coupling between the conduction band and the local moments goes quadratically in the electron density $x$.  
We conclude observing that interaction effects between conduction electrons in the B sublattice introduce the tendency to develop a finite out-of-plane ferromagnetic polarization. The analysis of the effect of interaction between conduction electrons is left to future studies.   

\subsubsection{HFL}

In the paramagnetic regime electrons are described by the mean-field Hamiltonian reads: 
\begin{equation}
\label{hfl_Hamiltonian}
    \bar H_{\rm mf}=\sum_{\bk\sigma}\xi_{\bk\sigma}c^\dagger_{\bk\sigma}c_{\bk\sigma}+\lambda\sum_\bk \chi^\dagger_\bk\chi_\bk+J_K\sum_{\bk\sigma}\left[V^*_\bk\Phi^*c^\dagger_\bk\chi_\bk+h.c.\right].
\end{equation}
We easily realize that the Green's function of the problem reads:
\begin{equation}
    \bm G^{-1}_\sigma(\bk,i\epsilon)=\left(\begin{matrix} i\epsilon-\lambda & -J_K\Phi V_\bk \\  -J_K\Phi^* V^*_\bk & i\epsilon-\xi_{\bk\sigma} \end{matrix}\right),
\end{equation}
so that the saddle-point equation for the bosonic amplitude $\Phi$ can be written as: 
\begin{equation}
\label{sce_phi}
    \Phi=-\frac{T}{2N}\sum_{\bk\sigma}\sum_{i\epsilon}V^*_\bk \mathcal G_{\chi,\sigma}(\bk,i\epsilon)J_K V_\bk \Phi G_{c,\sigma}(\bk,i\epsilon),
\end{equation}
where $\mathcal G_{\chi,\sigma}(\bk,i\epsilon)=(i\epsilon-\lambda)^{-1}$ is the bare Green's function and $G_{c,\sigma}(\bk,i\epsilon)=[i\epsilon-\xi_{\bk\sigma}-\Sigma_{c,\sigma}(\bk,i\epsilon)]^{-1}$ with 
\begin{equation}
\label{self_energy_c}
    \Sigma_{c,\sigma}(\bk,i\epsilon)=J^2_K|\Phi|^2|V_\bk|^2/(i\epsilon-\lambda).
\end{equation}
From the latter expression we readily find the quasiparticle residue: 
\begin{equation}
    Z_{\bk}=\frac{1}{1-\partial_z \Sigma_{c,\sigma}(\bk,z)|_{z=0}}=\frac{1}{1+J^2_K|\Phi|^2V^*_{\bk} V_{\bk}/\lambda^2}.
\end{equation}
The latter quantity evaluated at the Fermi surface of the heavy Fermi liquid gives the mass enhancement of the quasiparticles. 
Discarding the $\Phi=0$ solution the equation reduces to
\begin{equation}
    \frac{1}{J_K}=\frac{1}{2N}\sum_{\bk\sigma}\frac{f(E_{\bk-\sigma})-f(E_{\bk+\sigma})}{\sqrt{(\xi_{\bk\sigma}-\lambda)^2+4J^2_K|\Phi|^2|V_\bk|^2}}=\frac{1}{2N}\sum_{\bk\sigma}\frac{\theta(-E_{\bk-\sigma})}{\sqrt{(\xi_{\bk\sigma}-\lambda)^2+4J^2_K|\Phi|^2|V_\bk|^2}},
\end{equation}
where $E_{\bk\pm}=(\xi_{\bk\sigma}+\lambda)/2\pm\sqrt{(\xi_{\bk\sigma}-\lambda)^2+4J^2_K|\Phi|^2|V_\bk|^2}/2$ and the RHS is obtained taking the we took the zero temperature limit and considering the case $x<1$ where only the lower band is filled. In addition we also have the self-consistent equation for $\lambda$: 
\begin{equation}
\label{sce_lambda}
\frac{T}{N}\sum_{\bk\sigma}\sum_{i\epsilon}\frac{1}{i\epsilon-\lambda-\Sigma_{\chi,\sigma}(\bk,i\epsilon)}=1\implies\frac{1}{2N}\sum_{\bk\sigma}\theta(-E_{\bk-\sigma})\left(1+\frac{\xi_{\bk\sigma}-\lambda}{\sqrt{(\xi_{\bk\sigma}-\lambda)^2+4J^2_K|\Phi|^2|V_\bk|^2}}\right)=1.
\end{equation}
The onset of the HFL instability is determined looking at the instability condition of the normal state to interlayer hybridization. 
In this case the solution of Eq.~\eqn{sce_lambda} is $\lambda=0$, i.e. the local moments are pinned at the Fermi level, and Eq.~\eqn{sce_phi} becomes: 
\begin{equation}
\label{Kondo_critical}
    \frac{1}{J_K}+\Pi^0_{\chi c}(\bq\to0,i\Omega=0)=0,
\end{equation}
where $\Pi^0_{\chi c}(\bq\to0,\tau)=-\langle T_\tau(\hat V_{\bq=0}(\tau)\hat V^\dagger_{\bq=0})\rangle/N$, $\hat V_{\bq=0}=\sum_\bk V^*_\bk c^\dagger_\bk\chi_\bk /\sqrt{2}$ and $\Pi^0_{\chi c}(\bq\to0,i\Omega)=\int^\beta_0 e^{i\Omega\tau}\Pi^0_{\chi c}(\bq\to0,\tau)$. Expanding close to the bottom of the band Eq.~\eqn{Kondo_critical} becomes: 
\begin{equation}
    \frac{1}{J_K}-\frac{\rho_0}{2}\int^{E_\Lambda}_{\epsilon_F}d\epsilon\frac{|V_\epsilon|^2}{\epsilon-\epsilon_F}-\frac{\rho_0}{2}\int^{\epsilon_F}_{0}d\epsilon\frac{|V_\epsilon|^2}{\epsilon_F-\epsilon}=0,
\end{equation}
where $|V_\epsilon|=9a^2_M k^2_\epsilon/4$ with $k_\epsilon=\sqrt{2m_W\epsilon/\hbar^2}$.The integral is characterized by a log singularity at $\epsilon_F$. We introduce the IR cutoff $T_K$, $\epsilon\to\epsilon\pm T_K$, that regularize the divergence. Finally, by performing simple calculations we find: 
\begin{equation}
    T_K\simeq \epsilon_Fe^{-1/(\rho_0\bar J_K)}.
\end{equation}
Since the average over the Fermi surface $\bar J_K$ goes linearly with the doping $x$ in conduction band we find that the Kondo temperature is exponentially suppressed in the limit $x\to0^+$.
Finally, we notice that the Fermi energy is proportional to the filling factor in the WSe$_2$ layer, $\epsilon_F\propto x$. Away from the low-doping regime of exponential suppression we have $T_K\propto x$ which is different from the conventional result $T_K\propto \sqrt{x}$. The behavior $m/m^*\sim T_K\propto x$ is consistent with experimental results in Ref.~\cite{Zhao_2022}.

\section{Transport properties}
 
 In this section we detail the evaluation of the transverse and longitudinal conductivities in the various phases of the phase diagram. 
 
\subsection{Charge transport in the HFL}

\begin{figure}
    \centering
    \includegraphics[width=0.4\textwidth]{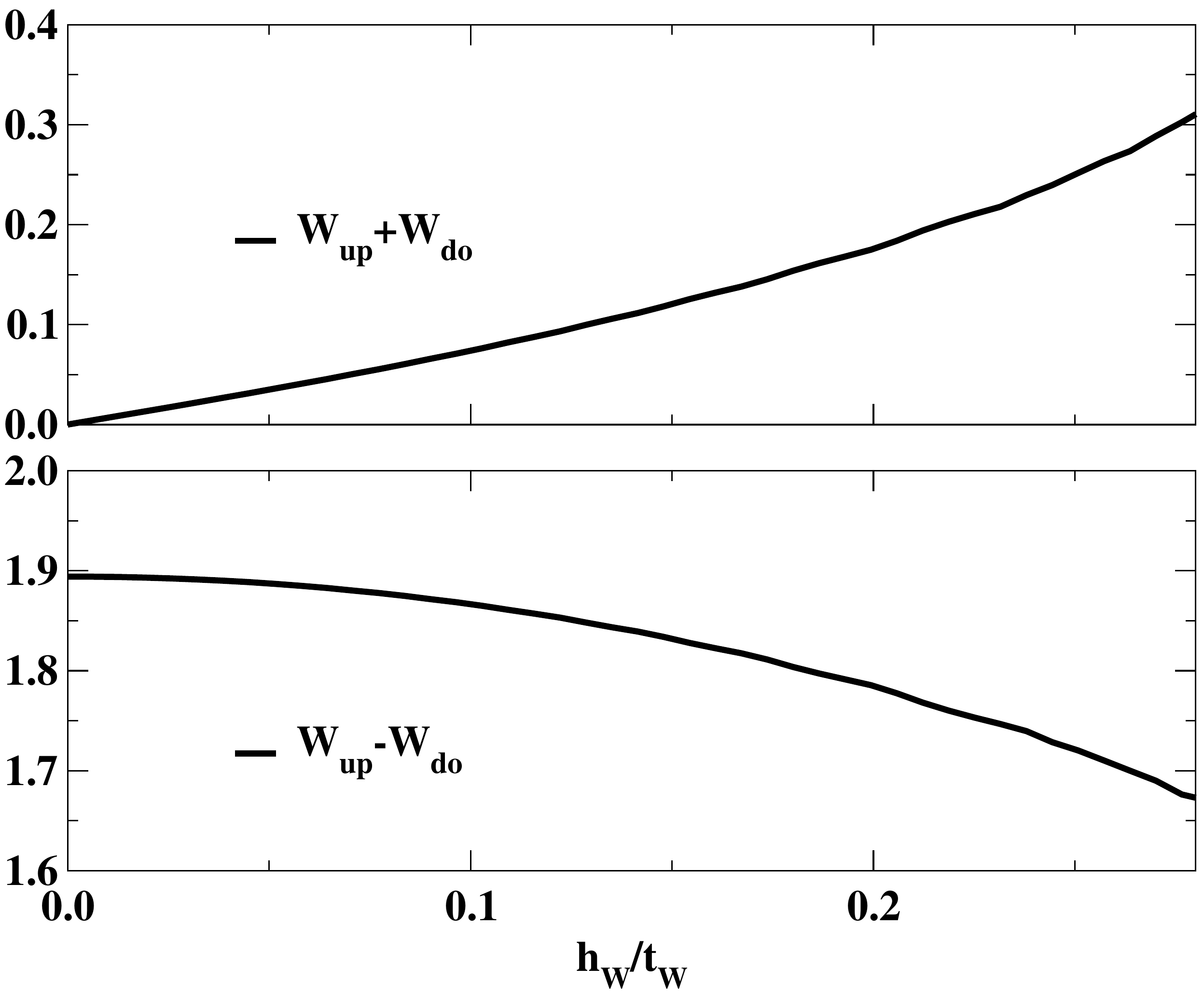}
    \caption{\textbf{Intrinsic topological response of the HFL.} Spin Hall effect (bottom panel) and anomalous Hall effect (top panel) in the HFL as a function of the Zeeman field. The calculation is performed at $J_K/t_W=1$, $x=0.65$. }
    \label{fig:sHE_aHE}
\end{figure}
\begin{figure}
    \centering
    \includegraphics[width=0.4\textwidth]{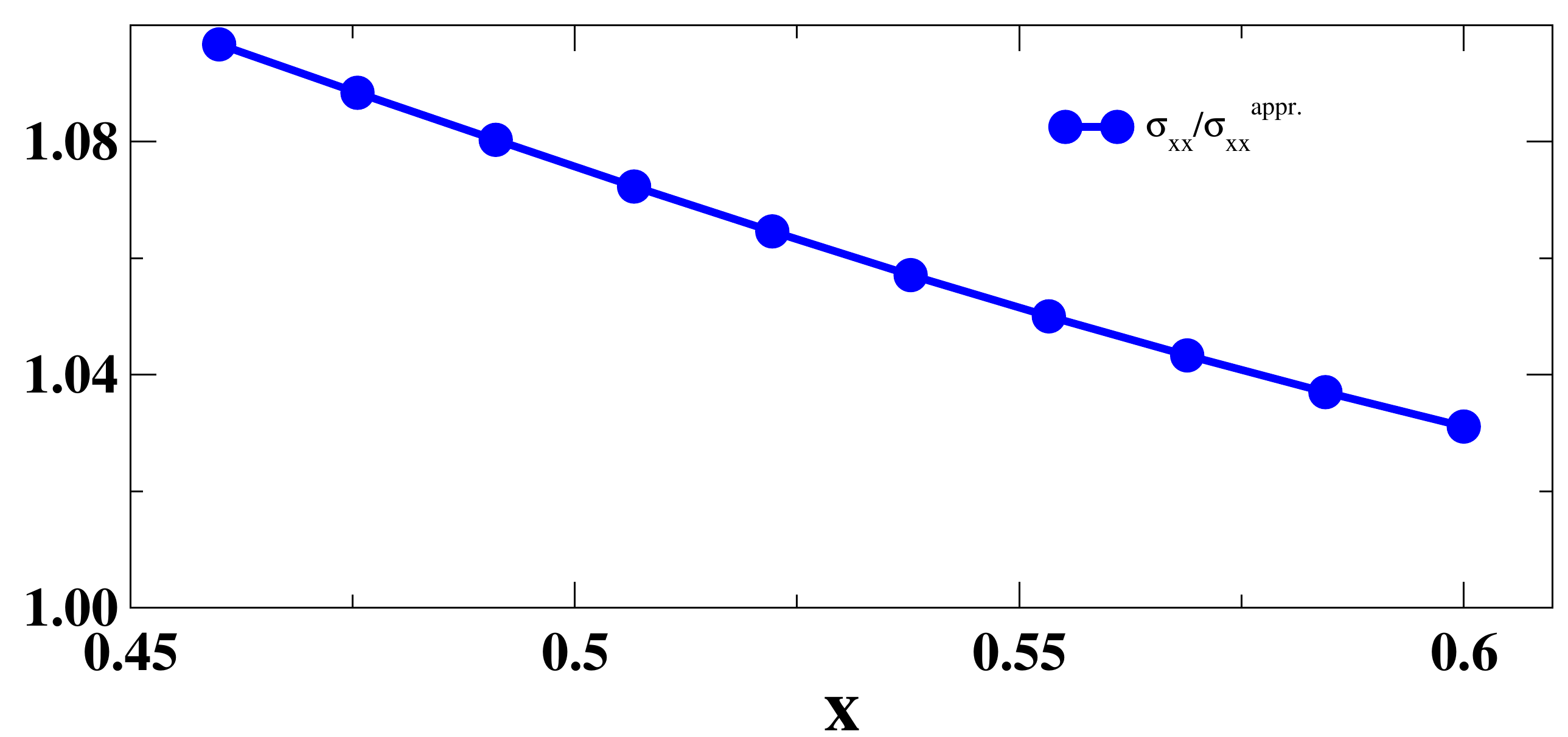}
    \caption{\textbf{Longitudinal conductivity.} Ratio between the numerical evaluation of Eq.~\eqn{longitudinal} and $\sigma_{xx}=e^2\tau(1-x)/\bar m^*$ obtained assuming a circular Fermi surface with average velocity.}
    \label{fig:circular}
\end{figure}

In the paramagnetic regime the local moments fractionalize giving rise to a finite density of holes in MoTe$_2$ layer. In this regime the field $\Phi_{\br}$ is equivalent to the holon operator carrying physical charge $-1$~\cite{Chowdhury_2018}, i.e. opposite to the electron charge. As a result both conduction electrons $c$ and spinon $\chi$ contribute to the charge current. Within the semiclassical Boltzmann equation approach and in the relaxation time approximation the transport properties are simply obtained as integrals over the quasiparticle Fermi surface~\cite{PhysRevB.43.193,Haldane_2004}: 
\begin{equation}
\label{longitudinal}
    \sigma_{xx}=\frac{3\sqrt{3} a^2_Me^2\tau}{8\pi^2}\sum_\sigma\oint_{FS} dk_t \frac{\bar v^x_{F,\sigma}\bar v^x_{F,\sigma}}{\hbar |\bar{\bm v}_{F,\sigma}|},
\end{equation}
and
\begin{equation}
\label{transverse}
    \sigma_{xy}\equiv \sigma^{\rm Ohm}_{xy}+\sigma^{\rm AH}_{xy}=\frac{3\sqrt{3} a^2_M}{8\pi^2}\frac{e^3\tau^2 B}{\hbar}\sum_\sigma\oint_{FS} dk_t \frac{\bar v^y_{F,\sigma}\left[\bar {v}^y_{F,\sigma}\partial_{k_x}-\bar {v}^x_{F,\sigma}\partial_{k_y}\right]\bar v^x_{F,\sigma}}{\hbar |\bar{\bm v}_{F,\sigma}|}+\frac{e^2}{2\pi h}\sum_\sigma\oint_{FS} dk_t\, \bm t\cdot\bm A_{\sigma}(\bk),
\end{equation}
where the anomalous Hall contribution $\sigma^{\rm AH}_{xy}$ arises from the circulation of the Berry connection $\bm A_{\sigma}(\bk)=i\braket{u_{\bk\sigma}}{\partial_{k_a} u_{\bk \sigma}}$, with $\ket{u_{\bk \sigma}}$ occupied eigenstate of the Hamiltonian in Eq.~\eqn{hfl_Hamiltonian}, along the FS. 
In Eqs.~\eqn{longitudinal} and~\eqn{transverse} $k_t$ is the component of $\bk$ along the tangent $\bm t$ to the FS curve. 
We notice that the FS of the heavy quasiparticle is obtained by the set of $\bk$ points solution of the equation:
\begin{equation}
    {\text{heavy FS}}: \xi_{\bk\sigma}-J^2_K|\Phi|^2 |V_\bk|^2/\lambda=0,
\end{equation}
where the second term comes from the conduction electron self-energy $\Sigma_{c,\sigma}(\bk,0)$ introduced in Eq.~\eqn{self_energy_c} computed at $i\omega=0$. The Fermi velocity $\bm v_{F,\sigma}$ is obtained replacing $\bk\to \bk+e\bA(t)/\hbar$ in $\bar\xi_{\bk\sigma}=Z_{\bk\sigma}[\xi_{\bk\sigma}-J^2_K|\Phi|^2 |V_\bk|^2/\lambda]$ and expanding to linear order around the FS we find $\bar \xi_{\bk_F+\frac{e}{\hbar}\bA(t)\sigma}\simeq e \bar{\bm v}_{F}\cdot\bA(t)$ where: 
\begin{equation}
    \bar{\bm v}_{F,\sigma} = \frac{Z_{\bk_F,\sigma}}{\hbar}\nabla_\bk[\xi_{\bk \sigma}-J^2_K|\Phi|^2 |V_\bk|^2/\lambda]\Big|_{\bk_F},
\end{equation}
and $Z_{\bk,\sigma}=[1-\partial_z\Sigma_{c,\sigma}(\bk,z)|_{z=0}]^{-1}$ is the quasiparticle weight. 
The evaluation of $\sigma_{xx}$ and $\sigma^{\rm Ohm}_{xy}$ is considerably simplified observing that the Fermi surface consists of a hole-pocket around $\bm\kappa'$ for spin $\uparrow$ and $\bm\kappa$ for spin $\downarrow$, respectively. Assuming a circular hole-like Fermi surface with average mass $\bar m^*$ the longitudinal contribution becomes 
\begin{equation}
\label{longitudinal_sigma}
    \sigma_{xx}=2\frac{e^2\tau}{\bar m^*}\left(\frac{\sqrt{3}a^2_M}{8\pi^2}\right)k^2_F\int^{2\pi}_0d\phi \cos^2\phi=\frac{e^2\tau}{\bar m^*}\left(2\pi k^2_F\frac{\sqrt{3}a^2_M}{8\pi^2}\right)=\frac{e^2\tau(1-x)}{\bar m^*}.
\end{equation}
As a sanity check we show in Fig.~\ref{fig:circular} the ratio between Eq.~\eqn{longitudinal_sigma} and the numerical evaluation of Eq.~\eqn{longitudinal} for various concentrations of electrons in the $W$-layer. By following the same line of reasoning we obtain $\sigma^{\rm Ohm}_{xy}=-e^3\tau^2 B(1-x)/\bar m^*{}^2$. We now look at the anomalous contribution $\sigma^{\rm AH}_{xy}$ which can be conveniently written as $\sigma^{\rm AH}_{xy}=e^2( W_\uparrow+ W_\downarrow)/h$ where $W_\sigma=\int d^2\bk  f(\epsilon_{\bk\sigma}) \Omega_{\sigma}(\bk)/(2\pi)$ and $\Omega_\sigma(\bk)=\partial_{k_x}A^y_\sigma(\bk)-\partial_{k_y}A^x_\sigma(\bk)$. Due to the opposite winding of spin $\uparrow$ and $\downarrow$ we find that in the absence of a magnetic field $W_\uparrow=- W_\downarrow$. We observe that the difference $ W_\uparrow- W_\downarrow$ gives a finite spin Hall (SH) conductivity~\cite{Kane_2005} even in the absence of the external field. Fig.~\ref{fig:sHE_aHE} shows $ W_\uparrow\pm  W_\downarrow$ as a function of the magnetic field at doping $x=0.46$ and $J_K/t_W=1$. The small value of $W_\uparrow+W_\downarrow$ at finite $B$ follows from the Berry curvature distribution which is peaked around the bare FS of the conduction electrons. As a result we find a small AH contribution and a large SH one.

\subsubsection{Expansion around the original Fermi surface: the topological Kondo Hamiltonian}

Here we derive the $\bk\cdot\bp$ Hamiltonian describing the regions around $\bm\kappa$ and $\bm\kappa'$ where the Fermi surface of spin $\uparrow$ and $\downarrow$, respectively, conduction electrons is located. The analysis clarifies the topological origin of the hybridization gap.
To start with we observe that the mean-field heavy Fermi liquid Hamiltonian reads: 
\begin{equation}
    H_{\sigma}(\bk)=\left(\begin{matrix} \lambda & J_K\Phi V_\bk \\  J_K\Phi^* V^*_\bk & \xi_{\bk\sigma} \end{matrix}\right).
\end{equation}
Expanding the form factor $V_{\bk}$ and the dispersion $\xi_{\bk \sigma}$ around $\bm\kappa$ we readily find:
\begin{equation}
\label{hfl_kp}
    H_{\uparrow}(\bm\kappa+\bk)=\left(\begin{matrix} \lambda & -\Delta_K(k_x-ik_y) \\  -\Delta_K(k_x+ik_y) & \hbar^2k^2/(2m_W)-\mu \end{matrix}\right),
\end{equation}
and $H_{\downarrow}(\bm\kappa'+\bk)=H^*_{\uparrow}(\bm\kappa-\bk)$. The crossing between the local moment $\chi$ and the $c$-electron dispersive band takes place of a circle with radius $k_c=\sqrt{2m_W(\mu+\lambda)}$ where the continuum model reduces to $H_{\uparrow}(\bm\kappa+\bk_c)=-\Delta_K k_c(e^{i\theta}\tau^-+e^{-i\theta}\tau^+)$. The interlayer Kondo hybridization lifts the degeneracy and induces winding in the two dimensional space associated with the interlayer degrees of freedom $\bm\tau$. The resulting Berry curvature can be readily obtained from the Kubo formula observing that the eigenvalues of Eq.~\eqn{hfl_kp} are described by the matrix $U(\bk)=\exp[-i\varphi_{\bk}(\bk\times\bm\tau)_z/2]$ with $\tan\varphi_{\bk}=\Delta_K k/d_z$ and $d_z$ projection of the Hamiltonian along $\tau^z$. The integral gives quantized Chern number $C_\uparrow=-C_\downarrow=1$. The model gives a topological Kondo metal which is adiabatically connected to a quantum Spin Hall Kondo insulator at filling $2$.

\subsection{The magnetic regime}
\begin{figure}
    \centering
    \includegraphics[width=0.6\textwidth]{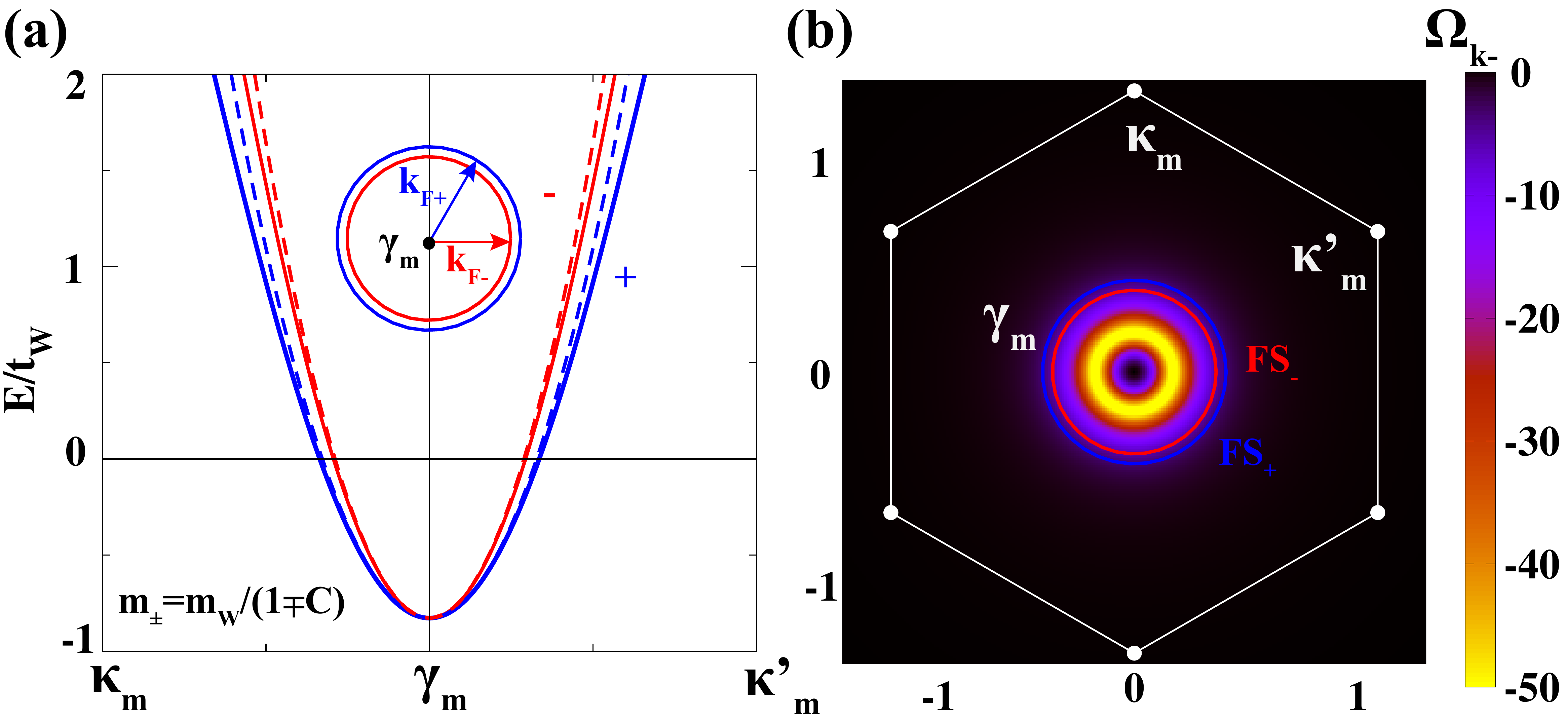}
    \caption{\textbf{Bandstructure and Berry curvature of the small Fermi surface magnetic state.} (a) Dispersion of the electrons close to the bottom of the conduction band. Dashed lines are obtained with the continuum Hamiltonian~\eqn{conduction_electrons} while the solid ones are the lowest energy bands of the mean-field Hamiltonian~\eqn{AFM_lattice}. (b) Berry curvature around the origin of the magnetic Brillouin zone. Solid lines show the two Fermi surfaces. The calculations has been performed at doping $x=0.08$, $J_K/t_W=1$, $J_H/t_W=0.025$ and magnetic field $h_W/t_W=0.08$.
    }
    \label{fig:fig_continuum}
\end{figure}
\begin{figure}
    \centering
    \includegraphics[width=0.45\textwidth]{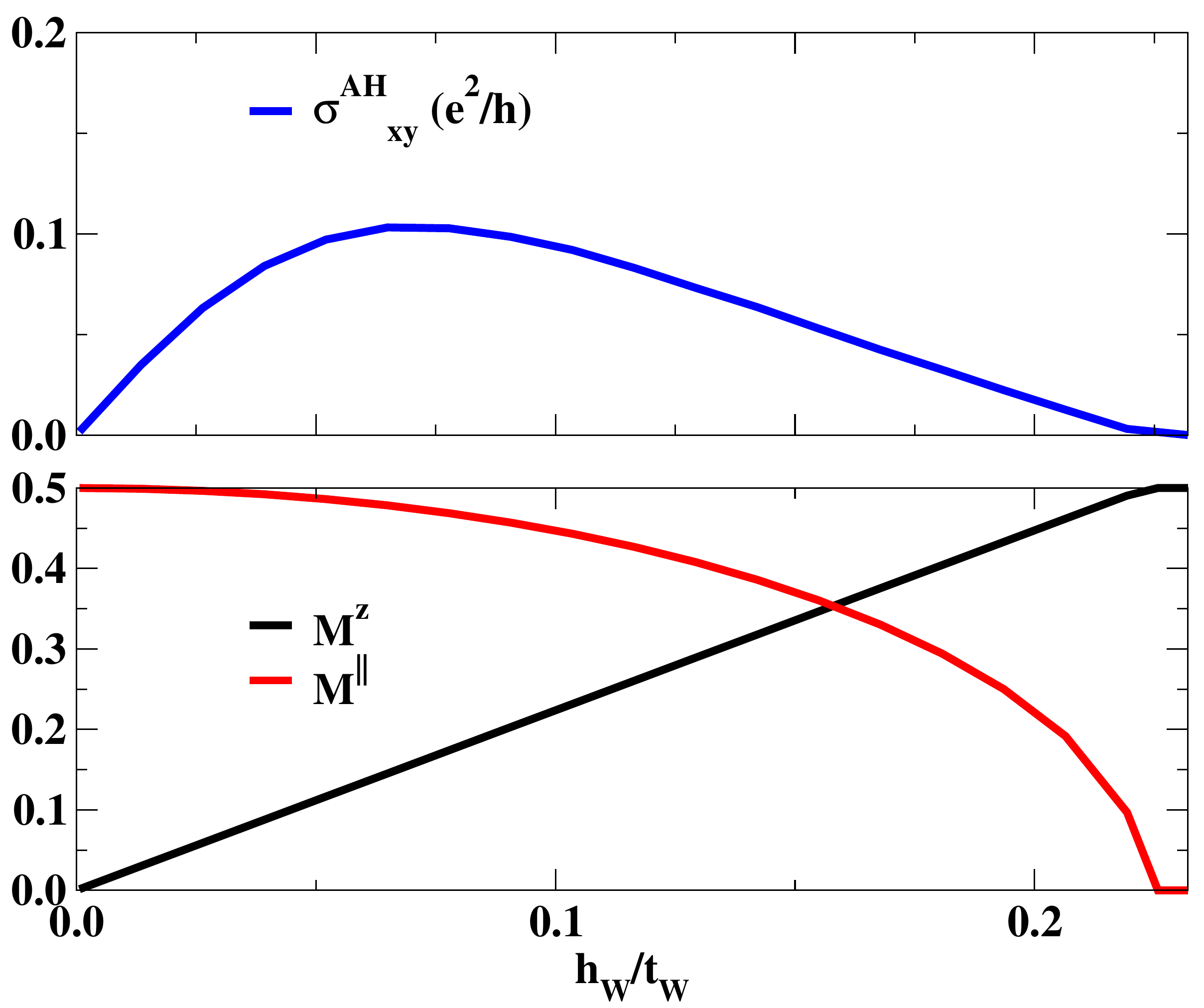}
    \caption{\textbf{Magnetic field in the small Fermi surface state.}  Bottom panel: evolution of the in-plane and out-of-plane magnetization as a function of the applied field. Top panel: anomalous Hall conductivity. The calculations has been performed at doping $x=0.05$, $J_K/t_W=1$, $J_H/t_W=0.05$. }
    \label{fig:ahe_m}
\end{figure}
\begin{figure}
    \centering
    \includegraphics[width=0.45\textwidth]{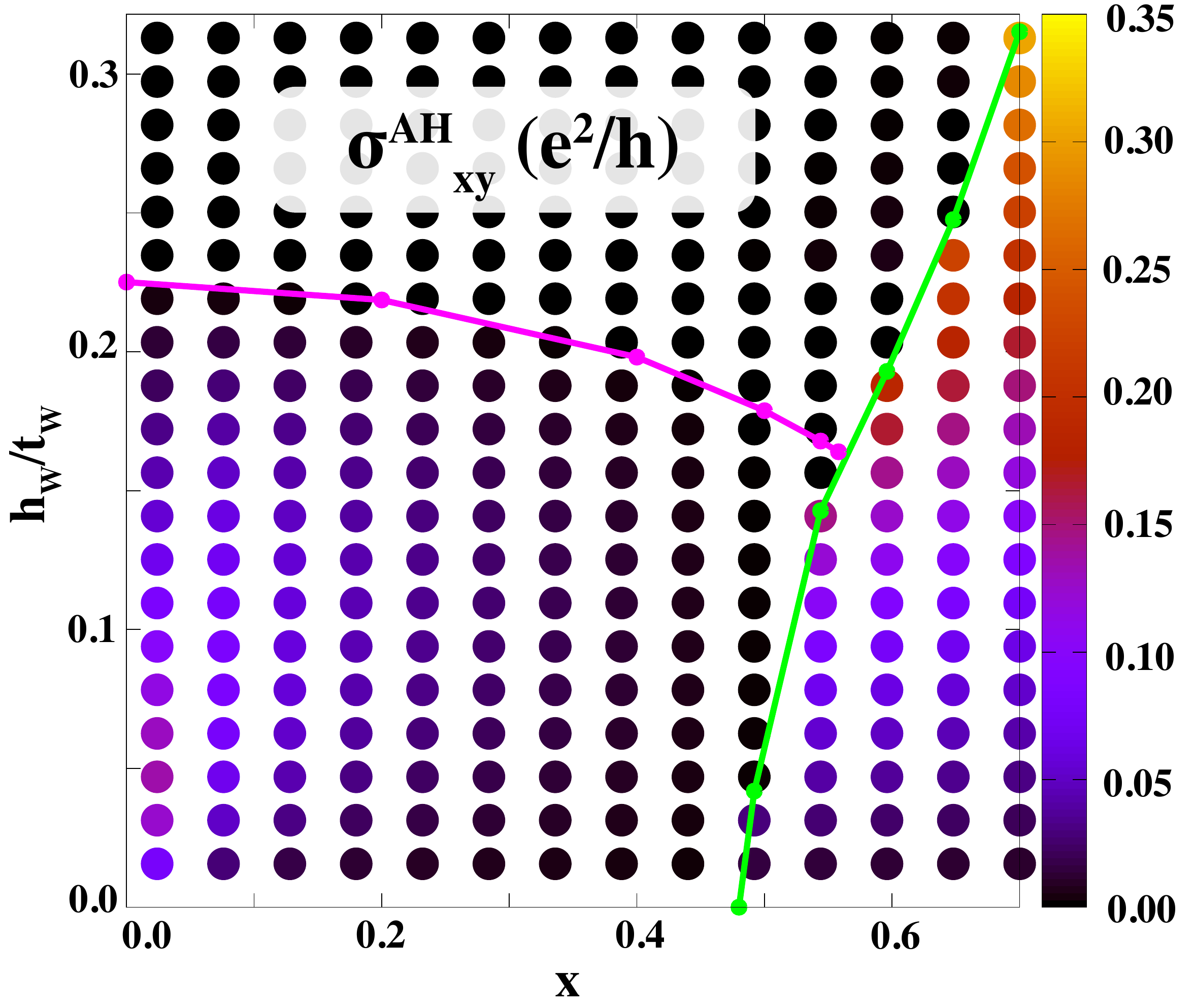}
    \caption{\textbf{Colormap of the AHE.} Intrinsic contribution to the anomalous Hall conductivity in the plane of $W$-layer concentration $x$ and $h_W$ Zeeman field.}
    \label{fig:ahe_2D}
\end{figure}

We now turn our attention in the magnetic regime where only the $c$-electrons contribute to charge transport. The low-energy Hamiltonian is given in Eq.~\eqn{conduction_electrons} and describes electrons with dispersion relation $\epsilon_{\bk\lambda}=\hbar^2 k^2/2m_\lambda$ shown in Fig.~\ref{fig:fig_continuum}(a) and group velocity $\bm v_{\bk\lambda}=\hbar \bk /m_\lambda$ with $m_\lambda=m_W/[1-\lambda J_K|M|/(4t_W)]$ and $\lambda=\pm$. Under the assumption of a single transport time, i.e. momentum- and band-independent, we apply Eqs.~\eqn{longitudinal} and \eqn{transverse} to find $\sigma_{xx}=e^2\tau x/\left(\sum_\lambda m_\lambda/2\right)$ and $\sigma^{\rm Ohm}_{xy}=e^3\tau^2 B x/\left(m_W\sum_\lambda m_\lambda/2\right)$. We now conclude our analysis considering the AHE contribution coming from the the Berry phase winding introduced by the $d_x(\bk)$ and $d_y(\bk)$ terms in Eq.~\eqn{conduction_electrons}. In the absence of an external magnetic field $M^z=0$ and the eigenstates are simply $\ket{u_{\bk\pm}}=(1,\pm e^{-2i \varphi_\bk})\sqrt{2}$ with $2\pi$-Berry phase around the origin of the magnetic Brillouin zone. A small Zeeman term opens a gap in the band structure and gives rise to a finite Berry curvature $\Omega_{\pm}(\bk)$:
\begin{equation}
\label{Berry_curvature}
    \Omega_{\pm}(\bk)=\pm\frac{64 h^2k^2\eta^2}{\left[16k^4\eta^2+\left(2h+\frac{m_{\uparrow}-m_{\downarrow}}{m_{\uparrow}m_{\downarrow}}k^2\right)^2\right]^{3/2}},
\end{equation}
where $\eta=9J_K M^\parallel/8$ and $m_{\uparrow/\downarrow}=m_W/[1\pm J_K M^z/(4t_W)]$.
We notice that the Berry curvature is an even function of $k$, vanishes quadratically at $k=0$ and takes its maximum value at finite momentum $k$. The maximum is located at $k^4_{\rm max}=h^2/(8\eta^2)$ for $M^z=0$. The momentum space distribution in the magnetic \mr Brillouin zone is given in Fig.~\ref{fig:fig_continuum}(b). We observe that from Eq.~\eqn{Berry_curvature} the anomalous Hall conductivity is obtained as: 
\begin{equation}
    \sigma^{\rm AH}_{xy}=\frac{e^2}{2\pi h}\sum_\lambda \int d^2\bk \Omega_{\lambda}(\bk) f(\epsilon_{\bk\lambda}-\mu),
\end{equation}
which in the zero temperature limit becomes: 
\begin{equation}
    \sigma^{\rm AH}_{xy}=\frac{e^2}{h}\int^{k^+_F}_{k^-_F} dk\,k \,\Omega_{+}(k)=\frac{e^2}{h}\frac{h[k^2(m_\downarrow-m_\uparrow)-2hm_\uparrow m_\downarrow]}{\sqrt{16k^4\eta^2(m_\uparrow m_\downarrow)^2+[k^2(m_\uparrow-m_\downarrow)+2hm_\uparrow m_\downarrow]^2}}\Bigg|^{k^+_F}_{k^-_F}.
\end{equation}
Fig.~\ref{fig:ahe_m} shows the evolution of the anomalous Hall conductivity in the low-density regime. 
Finally, Fig.~\ref{fig:ahe_2D} gives bird's-eye view over the anomalous Hall effect in the plane of the $W$-layer concentration and the Zeeman field.  

\end{document}